\documentclass[10pt]{article} % onecolumn (standard format)

\usepackage[utf8]{inputenc}
\usepackage{amsmath}
\usepackage{amsfonts}
\usepackage{amssymb}
\usepackage{mathtools}
\usepackage{listings}
\usepackage{mathrsfs}
\usepackage{times}
\usepackage{float}
\usepackage{color}
\usepackage{setspace}
\usepackage{color,soul}

\usepackage{amsmath,amsfonts,amsthm,eucal}
\usepackage{enumerate}
\usepackage[letterpaper,margin=3cm]{geometry}
\usepackage{float}
\usepackage[title]{appendix}
\usepackage{color}
\usepackage{tabularx}
\usepackage{siunitx}
\usepackage{empheq}
\usepackage[tableposition=below]{caption}
\usepackage[
pdfstartview=XYZ,
bookmarks=true,
colorlinks=true,
linkcolor=blue,
urlcolor=blue,
citecolor=blue,
pdftex,
bookmarks=true,
linktocpage=true,   % makes the page number as hyperlink in table of content
hyperindex=true
]{hyperref}

\usepackage{lipsum}
\usepackage[FIGBOTCAP,TABTOPCAP,bf,tight]{subfigure}

\usepackage{natbib}
\usepackage[pdftex]{graphicx}
\usepackage{epstopdf}
\usepackage{booktabs} 
\usepackage{tikz,mathpazo}
\usetikzlibrary{shapes.geometric, arrows}
\usepackage{caption}

\newcommand{\tensor}[1]{\ensuremath{\boldsymbol{#1}}}

\DeclareMathOperator{\grad}{\nabla}

\DeclareMathOperator{\diver}{\nabla\cdot}

\DeclareMathOperator{\tr}{tr}

\theoremstyle{remark}

\renewcommand{\vec}[1]{\ensuremath{\boldsymbol{#1}}}

\usepackage{algorithm}
\usepackage{algorithmicx}
\usepackage[noend]{algpseudocode}

\theoremstyle{definition}

\usepackage{longtable}
\usepackage{adjustbox}
\usepackage{framed}

\title{Multi-phase-field microporomechanics model for simulating ice lens growth and thaw in frozen soil} 

\begin{document}

\author{Hyoung Suk Suh\thanks{Department of Civil Engineering and Engineering Mechanics, 
 Columbia University, 
 New York, NY 10027.     \textit{h.suh@columbia.edu}  }       \and
        WaiChing Sun\thanks{Department of Civil Engineering and Engineering Mechanics, 
 Columbia University, 
 New York, NY 10027.
  \textit{wsun@columbia.edu}  (corresponding author)      %     \\
}
}

\maketitle

\begin{abstract}
This article presents a multi-phase-field poromechanics model that simulates the growth and thaw of ice lenses and the resultant frozen heave and thaw settlement in multi-constituent frozen soils. 
In this model, the growth of segregated ice inside the freezing-induced fracture is implicitly represented by the evolution of two phase fields that indicate the locations of segregated ice and the damaged zone, respectively. 
The evolution of two phase fields are driven by the driving forces that capture the physical mechanisms of ice and crack growths respectively, while the phase field governing equations are coupled with the balance laws such that the coupling among heat transfer, solid deformation, fluid diffusion, crack growth, and phase transition can be observed numerically. 
Unlike phenomenological approaches that indirectly captures the freezing influence on the shear strength, the multi-phase-field model introduces an immersed approach where both the homogeneous freezing and the ice lens growth are distinctively captured by the freezing characteristic function and the driving force accordingly. 
Verification and validation examples are provided to demonstrate the capacities of the proposed models. 
\end{abstract}

\section{Introduction}
Ice lens formation at the microscopic scale is a physical phenomenon critical for understanding the physics of frost heave and thawing settlement occurred at the field scale under the thermal cycles.
Since ice lens may affect the freeze-thaw action and cause frost heave and thawing settlement sensitive to the changing climate and environment conditions, knowledge on 
the mechanism for the ice lens growth is of practical value for many civil engineering applications in cold regions \citep{palmer2003frost, zhang2016analysis, li2017experimental, lake2017examining, ji2019frost}. 
For example, substantial heaving and settlement caused by the sequential formations and thawing of ice lenses lead to uneven deformation of the road which also damages the tires, suspension, and ball joints of vehicles. 
In the United States alone, it was estimated that two billion dollars had been spent annually to repair frost damage of roads \citep{dimillio1999quarter}. 
Moreover, extreme climate change over the last few decades have brought increasing attention to permafrost degradation, since unusual heat waves may cause weakening of foundations and increase the likelihood of landslides triggered by the abrupt melting of the ice lens \citep{nelson2001subsidence, nelson2002climate, streletskiy2012permafrost, leibman2014cryogenic, mithan2021topographic}. 
Under these circumstances, both the fundamental understanding of the ice lens growth mechanisms and the capacity to predict and simulate the effect beyond the one-dimensional models becomes increasingly important.

Since the pioneering work on the ice lens by Stephan Taber in the early 20th century \citep{taber1929frost, taber1930mechanics}, there has been a considerable amount of progress in the geophysics and fluid mechanics community to elucidate the mechanisms in the ice segregation process (e.g., \citep{peppin2013physics} and references cited therein). 
During the freezing phase, it is now known that the crystallized pore ice surrounded by a thin pre-melted water film develops a suction pressure (i.e., cryo-suction) that attracts the unfrozen water towards the freezing front \citep{wilen1995frost, dash1995premelting, dash2006physics}. 
These films remain unfrozen below the freezing temperature and form an interconnected flow network that supplies water to promote ice crystal growth. 
Accumulation of pore ice crystals accompanies the void expansion and micro-cracking of the host matrix, which may result in the formation of a horizontal lens of segregated ice. 
However, despite these substantial amounts of works, the criterion for the ice lens initiation and its detailed mechanism still remain unclear. 
Based on the thermo-hydraulic model proposed by Harlan \citep{harlan1973analysis}, Miller \citep{miller1972freezing, miller1977lens, o1985exploration} introduces a concept of stress partitioning and assumed that an ice lens starts to form if the solid skeleton experiences tensile stress. 
This idea has been further adopted and further generalized in \citep{fowler1989secondary, fowler1994generalized} via an asymptotic method. 
Gilpin \citep{gilpin1980model} suggests that the ice lens formation takes place when the ice pressure reaches the particle separation pressure depending on the particle size and the interfacial tension between the water and ice, whereas Zhou and Li \citep{zhou2012numerical} propose the idea of separation void ratio as a criterion for the ice lensing. 
Konrad and Morgenstern \citep{konrad1980mechanistic} present an alternative approach that can describe the formation and growth of a single ice lens based on segregation potential, of which the applicability has been demonstrated in \citep{nixon1982field, konrad19962, tiedje2012frost}. 
On the other hand, Rempel \citep{rempel2004premelting, rempel2007formation} develops regime diagrams that delineate the growth of a single lens, multiple lenses, and homogeneous freezing. 
In this line of work, the one-dimensional momentum and mass equilibrium equations are coupled with the heat flow in a step-freezing Stefan configuration to calculate the intermolecular force that drives the premelted fluid to the growing ice lenses. 
While the proposed method is helpful for estimating the lens thickness and spacing, the one-dimensional setting is understandably insufficient for the geo-engineering applications that require understanding on the implication of ice lenses on the shear strength. 
More recently, Style et al. \citep{style2011ice} propose a new theory on the ice lens nucleation by considering the cohesion of soil and the geometric supercooling of the unfrozen water in the pore space. 
Although the aforementioned studies formed the basis to shed light on explaining the ice lens formation, they are limited to the idealized one-dimensional problems and often idealized soil as a linear elastic material and hence not sufficient for applications that require a more precise understanding of the constitutive responses of the ice-rich soil.

Meanwhile, within the geomechanics and geotechnical engineering community, a number of theories and numerical modeling frameworks have been proposed based on the mixture theory and thermodynamics principles \citep{nishimura2009thm, zhou2013three, na2017computational, michalowski2006frost} with a variety of complexities and details. 
By adopting the premelting theory and considering the frozen soil as a continuum mixture of the solid, unfrozen water, and ice constituents, the freezing retention behavior of frozen soil can be modeled in a manner similar to those for the unsaturated soil. 
The resultant finite element implementation of these models enables us to simulate freeze-thaw effects in two- or three-dimensional spaces often with more realistic predictions on the solid constitutive responses. 
Nevertheless, the presence of crystal ices in the pores and that inside the expanded ice lens are often represented via phenomenological laws \citep{michalowski2006frost, ghoreishian2016constitutive}. 
Since the morphology, physics, and the mechanisms as well as the resultant mechanical characteristics of the ice lens and ice crystals in pores are profoundly different, it remains difficult to develop a predictive phenomenological constitutive law for an effective medium that represents the multi-constituent frozen soil with ice lenses \citep{wettlaufer2006premelting}.

This study is an attempt to reconcile the fluid mechanics and geotechnical engineering modeling efforts on modeling the frozen soil under changing climates. 
Our goal is to (1) extend the field theory for ice lens such that it is not restricted to one-dimensional problems and (2) introduce a framework that may incorporate more realistic path-dependent constitutive laws. 
As such, the coupling mechanism among phase transition, fluid diffusion, heat transfer, and solid mechanics can be captured without solely replying on phenomenological material laws. 
In particular, we introduce a mathematical framework and a corresponding finite element solver that may distinctively capture the physics of ice lens and freezing/thawing. 
We leverage the implicit representation of complex geometry afforded by a multi-phase-field framework to first overcome the difficulty on capturing the evolving geometry 
of the ice lens. 
By considering the ice lens as segregated bulk ice inside the freezing-induced fracture, we adopt two phase field variables that represent the state of the fluid phase constituent and the regularized crack topology, respectively. 
This treatment enables us to take account of the brittle fracture that may occur during ice lens growth and explicitly incorporate the addition and vanishing shear strength and bearing capacity of the ice lens under different environmental conditions. 
The phase transition of the fluid is modeled via the Allen-Cahn equation \citep{allen1979microscopic, boettinger2002phase}, while we adopt the phase field fracture framework to model brittle cracking in a solid matrix \citep{bourdin2008variational, miehe2010phase, borden2012phase}. 
The resultant framework may provide a fuller picture to analyzing the growth of the ice lens in the frozen soil, while verification exercises also confirm that the model may reduce to a classical thermo-hydro-mechanical model and isothermal poromechanics model under limited conditions.

The rest of the paper is organized as follows. 
Section \ref{sec:modeling_appr} summarizes the necessary ingredients for the mathematical framework, while we present the multi-phase-field microporomechanics model that describes the coupled behavior of a fluid-saturated phase-changing porous media in Section \ref{sec:microporomechanics}. 
For completeness, the details of the finite element formulation and the operator splitting solution strategy are discussed in Section \ref{sec:implementation}. 
Finally, numerical examples are given in Section \ref{sec:example} to verify, validate, and showcase the model capacity, which highlights its potential by simulating the growth and melting of multiple ice lenses.

As for notations and symbols, bold-faced and blackboard bold-faced letters denote tensors (including vectors which are rank-one tensors); 
the symbol '$\cdot$' denotes a single contraction of adjacent indices of two tensors 
(e.g.,\ $\vec{a} \cdot \vec{b} = a_{i}b_{i}$ or $\tensor{c} \cdot \tensor{d} = c_{ij}d_{jk}$); 
the symbol `:' denotes a double contraction of adjacent indices of tensor of rank two or higher
(e.g.,\ $\mathbb{C} : \vec{\varepsilon}$ = $C_{ijkl} \varepsilon_{kl}$); 
the symbol `$\otimes$' denotes a juxtaposition of two vectors 
(e.g.,\ $\vec{a} \otimes \vec{b} = a_{i}b_{j}$)
or two symmetric second-order tensors 
[e.g.,\ $(\tensor{\alpha} \otimes \tensor{\beta})_{ijkl} = \alpha_{ij}\beta_{kl}$]. 
We also define identity tensors: $\tensor{I} = \delta_{ij}$, $\mathbb{I} = \delta_{ik}\delta_{jl}$, and $\bar{\mathbb{I}} = \delta_{il}\delta_{jk}$, where $\delta_{ij}$ is the Kronecker delta.
As for sign conventions, unless specified, the directions of the tensile stress and dilative pressure are considered as positive.

%\section{Modeling approaches}
\section{Kinematics and effective stress principle for frozen soil with ice lens}
\label{sec:modeling_appr}
In this section, we introduce the ingredients necessary to derive the field theory for the phase field modeling of frozen soil presented later in Section \ref{sec:microporomechanics}. 
Similar to the treatments in \citep{nishimura2009thm}, \citep{zhou2013three}, and \citep{na2017computational}, we first assume that the frozen soil is fully saturated with either water or ice and therefore idealize the frozen soil as a three-phase continuum mixture that consists of solid, water, and ice phase constituents whereas the ice lens is a special case in which the solid skeleton no longer holds 
bearing capacity. 
This treatment enables us to formulate a multi-phase-field approach to employ two phase field variables as indicator functions for the state of the pore fluid (in ice or water form) \citep{warren1995prediction, boettinger2002phase, sweidan2020unified} and that of the solid skeleton (in damage or intact form) \citep{bourdin2008variational, miehe2010phase, borden2012phase}. 
We then extend the effective stress theory originated from damage mechanics \citep{chaboche1988continuum} to incorporate the internal stress of ice lenses caused by the deformation of the effective medium into the Bishop's effective stress principle for frozen soil where the introduction of phase field provide smooth transition of the material states for both the pore fluid and the solid skeleton. 
This procedure allows us to incorporate both the capillary pressure of the ice crystal surrounded by the water thin film as well as the volumetric and deviatoric stresses triggered by the deformation of ice lens.

\subsection{Continuum representation and kinematics}
\label{sec:con_rep}
Based on the mixture theory, we idealize our target material as a multiphase continuum where the solid, water, and ice phase constituents are overlapped. 
For simplicity, this study assumes that there is no gas phase inside the pore such that the pore space is either occupied by water or ice. 
The volume fractions of each phase constituent are defined as,
\begin{equation}
\label{eq:vol_frac}
\phi^s = \frac{d V_s}{d V}
\: \: ; \: \:
\phi^w = \frac{d V_w}{d V}
\: \: ; \: \:
\phi^i = \frac{d V_i}{d V}
\: \: ; \: \:
\phi^s + \displaystyle\sum_{\alpha =\lbrace w, i \rbrace} \phi^{\alpha} = 1,
\end{equation}
where the indices $s$, $w$, and $i$ refer to the solid, water, and ice phase constituents, respectively, while $dV = dV_s + dV_w + dV_i$ denote the total elementary volume of the mixture. 
Note that an index used as a subscript indicates the intrinsic property of a phase constituent, while it is used as a superscript when referring to a partial property of the entire mixture. 
By letting $\rho_s$, $\rho_w$, and $\rho_i$ denote the intrinsic mass densities of the solid, water, and ice, respectively, the partial mass densities of each phase constituent are given by, 
\begin{equation}
\label{eq:partial_density}
\rho^s = \phi^s \rho_s
\: \: ; \: \:
\rho^w = \phi^w \rho_w
\: \: ; \: \:
\rho^i = \phi^i \rho_i
\: \: ; \: \:
\rho^s + \displaystyle\sum_{\alpha =\lbrace w, i \rbrace} \rho^{\alpha} = \rho,
\end{equation}
where $\rho$ is the total mass density of the entire mixture. 
We also define the saturation ratios for the fluid phase constituents $\alpha = \lbrace w, i \rbrace$ as:
\begin{equation}
\label{eq:saturation}
S^w = \frac{\phi^w}{\phi}
\: \: ; \: \:
S^i = \frac{\phi^i}{\phi}
\: \: ; \: \:
\displaystyle\sum_{\alpha =\lbrace w, i \rbrace} S^{\alpha} = 1,
\end{equation}
where $\phi = 1 - \phi^s$ is the porosity.

Since the solid ($s$), water ($w$), and ice ($i$) phases do not necessarily follow the same trajectory, each constituent possesses its own Lagrangian motion function that maps the position vector of the current configuration $\vec{x}$ at time $t$ to their reference configurations. 
In this study, we adopt a kinematic description that traces the motion of the solid matrix by following the classical theory of porous media \citep{bowen1980incompressible, zienkiewicz1999computational, ehlers2002foundations, coussy2004poromechanics}. 
Hence, the motion of the solid phase is described by using the Lagrangian approach via its displacement vector $\vec{u}(\vec{x},t)$, whereas the fluid phase ($\alpha = \lbrace w, i \rbrace$) motions are described by the modified Eulerian approach via relative velocities $\tilde{\vec{v}}_w$ and $\tilde{\vec{v}}_i$, instead of their own velocity fields $\vec{v}_w$ and $\vec{v}_i$, i.e.,
\begin{equation}
\label{eq:rel_vec}
\tilde{\vec{v}}_{\alpha} = \vec{v}_{\alpha} - \vec{v},
\end{equation}
where $\vec{v} = \dot{\vec{u}}$ is the solid velocity, while $\dot{( \bullet )} = \mathrm{d} (\bullet) / \mathrm{d} t$ is the total time derivative following the solid matrix.

\subsection{Multi-phase-field approximation of freezing-induced crack}
\label{sec:multi_phase_field}
\begin{figure}[h]
\centering
\includegraphics[height=0.25\textwidth]{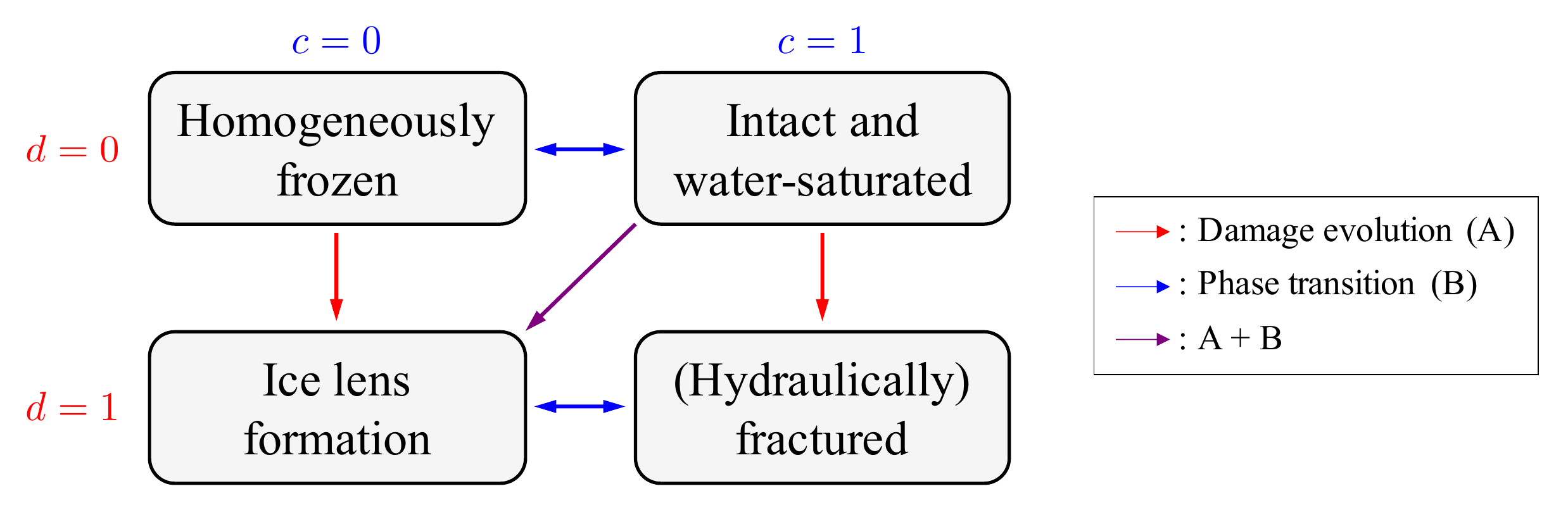}
\caption{Schematic of multi-phase-field approach coupled with a thermo-hydro-mechanical model.}
\label{fig:multi_phase_field}
\end{figure}

In this current study, we assume that the path-dependent constitutive responses of the frozen soil is due to the fracture in the brittle regime and the growth/thaw of the ice lens in the void space that could be opened by the expanded ice. While plasticity of the solid skeleton as well as the damage and creeping of the segregated ice may also play important roles on the mechanisms 
of the frost heave and thaw settlement, they are out of the scope of this study. 
As such, this study follows Miller's theory which assumes that a new ice lens may only form if and only if the compressive effective stress becomes zero or negative \citep{miller1972freezing, miller1977lens, o1982numerical, o1985exploration}. 
Since opening up the void space is a necessary condition for the ice lens to grow inside, we introduce a phase field model that captures
the crack growth potentially caused by the ice lenses growth.  
In this work, our strategy is to adopt diffuse approximations for both the phase transition of the pore fluid and the crack topology, where each requires a distinct phase field variable. 
As illustrated in Fig.~\ref{fig:multi_phase_field}, introducing two phase fields not only enables us to distinguish the homogeneous freezing from the ice lens growth but also leads to a framework that can be considered as a generalization of a thermo-hydro-mechanical model. 

The first phase field variable $c \in [0, 1]$ used in this study is an order parameter that models the freezing of water (melting of ice) in a regularized manner \citep{warren1995prediction, sweidan2020unified}. 
In other words, we employ a diffuse representation of the ice-water interface using variable $c$ that is a function of $\vec{x}$ and $t$:
\begin{equation}
\label{eq:phase_field_c}
c = c(\vec{x}, t) \text{ with } 
\begin{dcases}
c = 0 & \text{: completely frozen}, \\
c = 1 & \text{: completely unfrozen}, \\
c \in (0, 1) & \text{: diffuse ice-water interface}, 
\end{dcases}
\end{equation}
which is the solution of the Allen-Cahn phase field equation \citep{allen1979microscopic, boettinger2002phase} that will be presented later in Section \ref{sec:heat_and_phase_change}. 
Based on this setting, we consider the degree of saturation of water as an interpolation function of the phase field $c$, i.e., $S^w = S^w(c)$, which monotonically increases from 0 to 1 as,
\begin{equation}
\label{eq:S_w}
S^w(c) = c^3 (10 - 15c + 6c^2).
\end{equation}
Note that the evolution of the phase field variable $c$ itself does not necessarily imply the ice lens growth since both the homogeneously frozen region and segregated ice can reach $c = 0$, regardless of the level of the effective stress or stored energy that drives the crack growth (Fig. \ref{fig:multi_phase_field}).

The second phase field variable $d \in [0, 1]$ adopted in this study is a damage parameter that treats the sharp discontinuity as a diffusive crack via implicit function \citep{bourdin2008variational, miehe2010phase, borden2012phase, suh2020phase}. 
In particular, we have:
\begin{equation}
\label{eq:phase_field_d}
d = d(\vec{x}, t) \text{ with } 
\begin{dcases}
d = 0 & \text{: intact}, \\
d = 1 & \text{: damaged}, \\
d \in (0,1) & \text{: transition zone}, 
\end{dcases}
\end{equation}
to approximate the fracture surface area $A_{\Gamma}$ as $A_{\Gamma_d}$, which is the volume integration of crack surface density $\Gamma_d(d, \grad{d})$ over a body $\mathcal{B}$, i.e.,
\begin{equation}
\label{eq:crack_surf_density}
A_{\Gamma} \approx A_{\Gamma_d}
=
\int_{\mathcal{B}} \Gamma_d(d, \grad{d}) \: dV
\: \: ; \: \:
\Gamma_d(d, \grad{d}) = \frac{d^2}{2 l_d} + \frac{l_d}{2} ( \grad{d} \cdot \grad{d} ),
\end{equation} 
where $l_d$ is the length scale parameter that controls the size of the transition zone. 
In this case, the crack resistance force $\mathcal{R}_d$ can be expressed as,
\begin{equation}
\label{eq:crack_resistance}
\mathcal{R}_d = \frac{\partial W_d}{\partial d} - \diver{\left( \frac{\partial W_d}{\partial \grad{d}} \right)}
\: \: ; \: \:
W_d = \mathcal{G}_d \Gamma_d \left( d, \grad{d} \right),
\end{equation}
where $\mathcal{G}_d$ is the critical energy release rate that quantifies the resistance to cracking. 
As hinted in Fig. \ref{fig:multi_phase_field}, in order to guarantee crack irreversibility, the thermodynamic restriction $\dot{\Gamma}_d \ge 0$ must be satisfied \citep{miehe2010phase, miehe2010thermodynamically, choo2018cracking, heider2021review} unlike the reversible freezing and thawing process. 
A necessary condition for thermodynamic consistency is to adopt a quadratic stiffness degradation function $g_d(d) = (1-d)^2$ which satisfies the following conditions:
\begin{equation}
\label{eq:degradation}
g_d(0) = 1
\: \: ; \: \:
g_d(1) = 0
\: \: ; \: \:
\frac{\partial g_d(d)}{\partial d} \le 0 \text{ for } d \in [0, 1].
\end{equation}
Based on this setting, we define an indicator function $\chi^i \in [0,1]$ for the segregated ice inside the freezing-induced fracture as follows:
\begin{equation}
\label{eq:ice_lens_indicator}
\chi^i(c,d) = [1 - S^w(c)][1 - g_d(d)],
\end{equation}
such that $\chi^i = 1$ implies the formation of the ice lens, which is different from the in-pore crystallization of the ice phase constituent.

\subsection{Effective stress principle}
\label{sec:eff_stress}
Leveraging the similarities between freezing/thawing and drying/wetting processes, Miller and co-workers \citep{miller1972freezing, miller1977lens, o1982numerical, o1985exploration} propose the concept of neutral stress that partitions the net pore pressure $\bar{p}$ into the pore water and pore ice pressures ($p_w$ and $p_i$), respectively:
\begin{equation}
\label{eq:pore_pressure_partitioning}
\bar{p} = S^w(c) p_w + [1 - S^w(c)] p_i.
\end{equation}
Clearly, Eq.~\eqref{eq:pore_pressure_partitioning} alone cannot capture the deviatoric stress induced by the deformation of the ice lens. 
Previous efforts on modeling frozen soil often relies on a extension of critical state theory that evolves the yield function according to the degree of saturation of ice (and therefore introduces the dependence of the tensile and shear strength on the presence of ice) \citep{nishimura2009thm, na2017computational}. 
However, this treatment is not sufficient to consider the soil that may become brittle at low temperature due to the low moisture content and the influence of ice lens on the elasticity. 
Hence, this study extends Miller's approach into a phase field framework by decomposing the effective stress tensor $\bar{\tensor{\sigma}}'$ into two partial stresses for the solid and ice lens via the damage phase field doubled as a weighting function, i.e., 
\begin{equation}
\label{eq:effective_stress_partitioning}
\bar{\tensor{\sigma}}' = g_d(d) \tensor{\sigma}'_{\text{int}} + [1 - g_d(d)] \tensor{\sigma}'_{\text{dam}}.
\end{equation}
where the second term on the right hand side of Eq. \eqref{eq:effective_stress_partitioning} depends on the saturation $S^w(c)$. 
Specifically, the effective stress contribution from the solid skeleton $\tensor{\sigma}'_{\text{int}}$ degrades due to the damage when ice lens grows, but may also evolve by the change of $\tensor{\sigma}'_{\text{dam}}$ in the presence of ice lens [for instance, see Eq. \eqref{eq:constitutive_solid} in Section \ref{sec:freezing_frac}]. 
Similar models that capture the constituent responses of porous media consisting of multiple solid constituents can also be found in \citep{borja2020cam}. 
In addition, this study also considers the volumetric expansion due to the phase transition from water to ice while neglecting the thermal expansion or contraction of each phase constituent. 
Hence, the total stress of the mixture $\tensor{\sigma}$ can be expressed as, 
\begin{equation}
\label{eq:total_stress}
\tensor{\sigma} = \bar{\tensor{\sigma}}' - \bar{p} \tensor{I} - \phi [1-S^w(c)] \bar{\alpha}_v K_i \tensor{I},
\end{equation}
where $\bar{\alpha}_v = g_d(d) \alpha_{v,\text{int}} + [1 - g_d(d)] \alpha_{v,\text{dam}}$ is the net volumetric expansion coefficient which is influenced by the evolution of the fracture. 
In particular, we assume that the volumetric expansion coefficient of the ice lens $\alpha_{v,\text{dam}}$ is greater than that of the pore ice crystal $\alpha_{v,\text{int}}$ due to the degradation of the solid skeleton.

\section{Multi-phase-field microporomechanics model for phase-changing porous media}
\label{sec:microporomechanics}
This section presents the balance principles and constitutive laws that capture the thermo-hydro-mechanical behavior of the phase-changing porous media. 
We first introduce the coupled field equations that govern the heat transfer and the ice-water phase transition processes which involve the latent heat effect. 
Unlike previous studies that model the phase transition of the pore fluid by using the semi-empirical approach which links either the Gibbs-Thomson equation \citep{zhou2013three} or the Clausius-Clapeyron equation \citep{nishimura2009thm, na2017computational} with the van Genuchten curve \citep{van1980closed}, we adopt the Allen-Cahn type phase field model \citep{allen1979microscopic, boettinger2002phase} with a driving force that depends both on the temperature and the damage. 
We then present microporomechanics and phase field fracture models that complete the set of governing equations, which is not only capable of simulating freeze-thaw action but also the freezing-induced or hydraulically-driven fractures. 
The implications of our model will be examined via numerical examples in Section \ref{sec:example}.

\subsection{Thermally induced phase transition}
\label{sec:heat_and_phase_change}
\subsubsection{Heat transfer}
\label{sec:heat_trans}
Since underground freezing and thawing processes may span over long temporal scales, this study employs a single temperature field $\theta$ by assuming that all the phase constituents reach a local thermal equilibrium instantly \citep{suh2021asynchronous}. 
We also neglect thermal convection by considering the case where the target material possesses low permeability. 
Let $e$ denote the internal energy per unit volume and $\vec{q}$ the heat flux. 
Then, the energy balance of the entire mixture can be expressed as \citep{gelet2012thermo, suh2021asynchronous}, 
\begin{equation}
\label{eq:en_bal}
\dot{e} = - \diver{\vec{q}} + \hat{r},
\: \: ; \: \:
e = e^s + \displaystyle\sum_{\alpha =\lbrace w, i \rbrace} e^{\alpha},
\end{equation}
where $\hat{r}$ indicates the heat source/sink, $e^s = \rho^s c_s \theta$ and $e^{\alpha} = \rho^{\alpha} c_{\alpha} \theta$ are the partial energies for the solid and fluid phase constituents, respectively, while $c_s$ and $c_{\alpha}$ are their heat capacities. 
Assuming that the freezing temperature of water (i.e., melting temperature of ice) remains constant: $\theta_m = 273.15$ K, the internal energy of the entire mixture $e$ in Eq.~\eqref{eq:en_bal} can be rewritten as, 
\begin{equation}
\label{eq:internal_energy}
e = \rho^s c_s \theta + (\rho^w c_w + \rho^i c_i) (\theta - \theta_m) + (\rho^w c_w + \rho^i c_i ) \theta_m.
\end{equation}
From the relations shown in Eqs.~\eqref{eq:vol_frac}-\eqref{eq:saturation}, substituting Eq.~\eqref{eq:internal_energy} into Eq.~\eqref{eq:en_bal} yields the following:
\begin{equation}
\label{eq:en_bal2}
( \rho^s c_s + \rho^w c_w + \rho^i c_i ) \dot{\theta}
+ \phi \left[ (\rho_w c_w - \rho_i c_i)(\theta - \theta_m) + \rho_i L_{\theta} \right] \dot{S}^w(c)
+ \diver{\vec{q}} = \hat{r},
\end{equation}
where:
\begin{equation}
\label{eq:lat_heat}
L_{\theta} = \left( \frac{\rho_w}{\rho_i} c_w - c_i \right) \theta_m
\end{equation}
is the latent heat of fusion which is set to be $L_{\theta} = 334$ kJ/kg for pure water \citep{warren1995prediction, loginova2001phase, nishimura2009thm, alexiades2018mathematical}. 
Notice that the second term on the left-hand side of Eq.~\eqref{eq:en_bal2} describes the energy associated with the phase change of the fluid phase constituent $\alpha = \lbrace w, i \rbrace$, which is responsible for the constant temperature during the transformation processes, i.e., where $c$ is changing with time since $\dot{S}^w(c) = \lbrace \partial S^w(c) / \partial c \rbrace \dot{c}$.  
For the constitutive model that describes the heat conduction, this study adopts Fourier's law where the heat flux can be written as the dot product between the effective thermal conductivity and the temperature gradient, i.e.,
\begin{equation}
\label{eq:fourier}
\vec{q} = - \left( \phi^s \kappa_s + \displaystyle\sum_{\alpha =\lbrace w, i \rbrace} \phi^{\alpha} \kappa_{\alpha} \right) \cdot \grad{\theta},
\end{equation}
where $\kappa_s$ and $\kappa_{\alpha}$ denote the intrinsic thermal conductivities of the solid and fluid phase constituents, respectively. 
This volume-averaged approach, however, is only valid for the case where all the phase constituents are connected in parallel. 
Although there exists alternative homogenization approaches such as Eshelby's equivalent inclusion method \citep{eshelby1957determination, hiroshi1986equivalent, sun2015stabilized}, determination of correct effective thermal conductivity often requires knowledge of the pore geometry and topology \citep{hiroshi1986equivalent, lee2017particle,  suh2018modification}. 
Since the information is not always readily approachable, this extension will be considered in the future.

\subsubsection{Phase transition}
\label{sec:phase_trans}
\begin{figure}[h]
\centering
\subfigure[]{\label{fig:g_c}\includegraphics[height=0.375\textwidth]{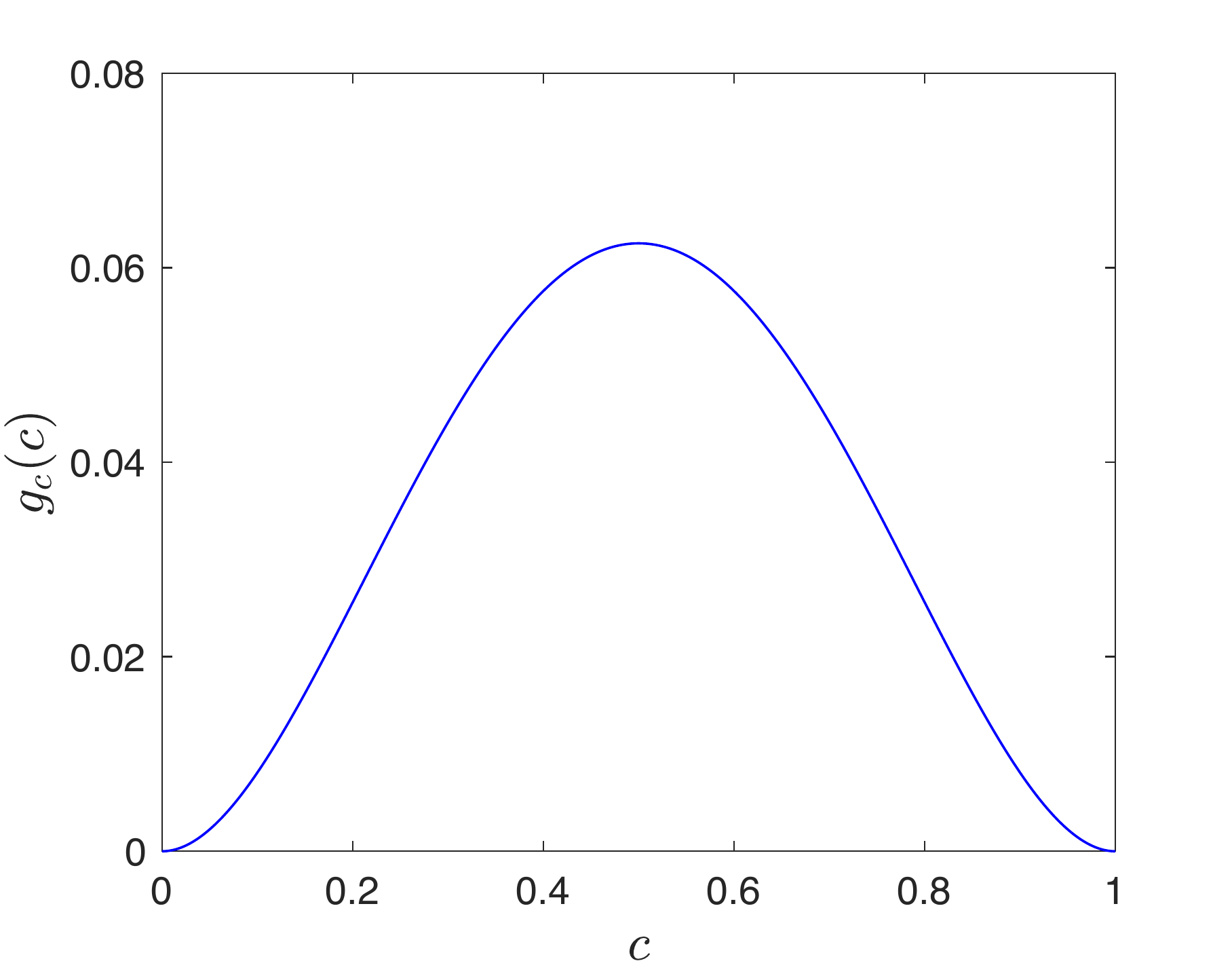}}
\hspace{0.01\textwidth}
\subfigure[]{\label{fig:p_c}\includegraphics[height=0.375\textwidth]{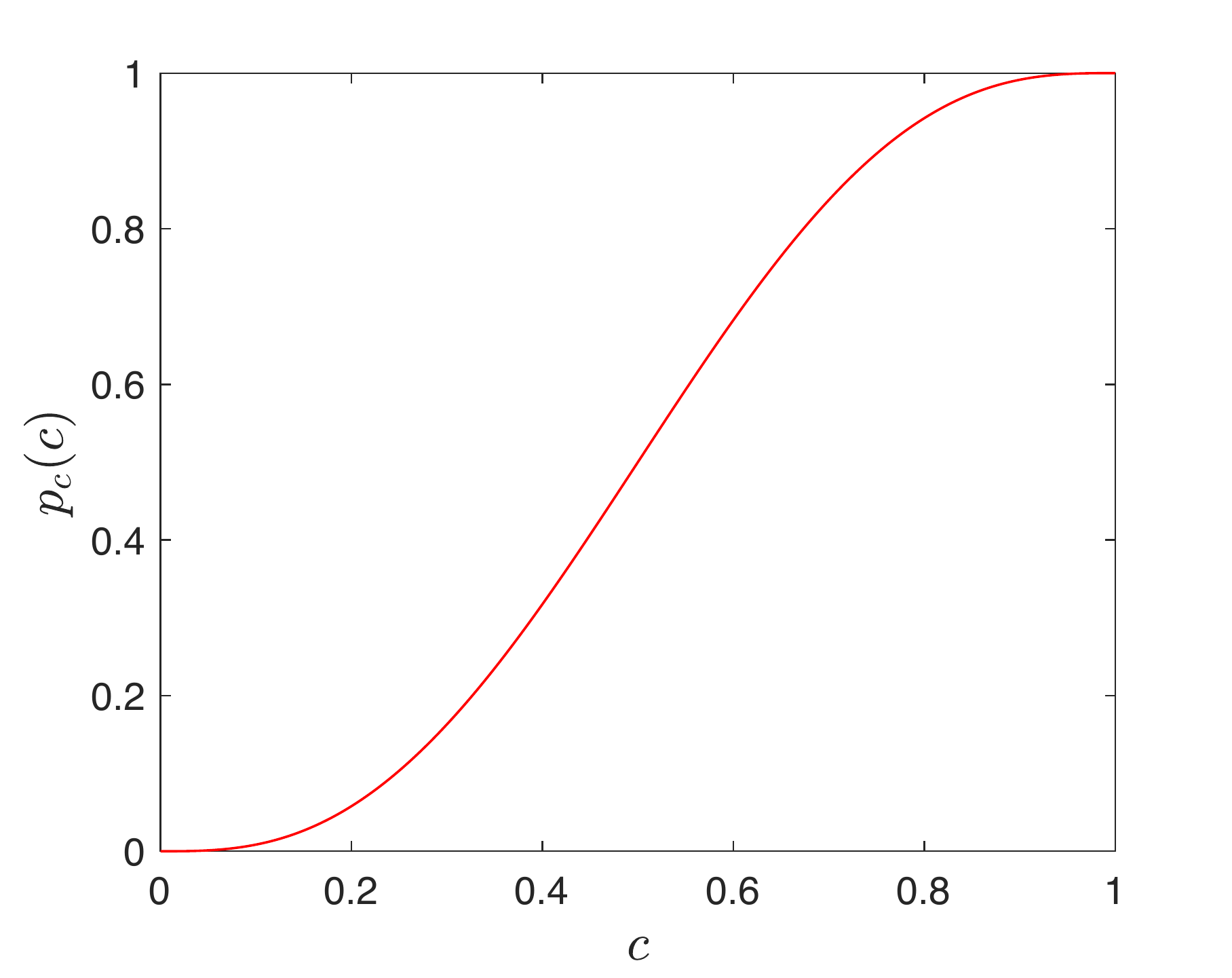}}
\caption{(a) The double well potential $g_c(c)$, and (b) the interpolation $p_c(c)$ functions.}
\label{fig:driving_force_plot}
\end{figure}

By using the phase field variable $c$ defined in Eq.~\eqref{eq:phase_field_c}, we adopt the Allen-Cahn model that is often used to simulate dendrite growth or multi-phase flow \citep{allen1979microscopic, takaki2014phase, aihara2019multi}. 
Following \citep{boettinger2002phase}, we consider one of the simplest forms of the Gibbs free energy functional $\Psi_c$: 
\begin{equation}
\label{eq:free_en}
\Psi_c 
= \int_{\mathcal{B}} \psi_c \: dV 
= \int_{\mathcal{B}} f_c(\theta, c) + \frac{\epsilon_c^2}{2} | \grad{c} |^2 \: dV,
\end{equation}
where $f_c(\theta, c)$ is the free energy density that couples the heat transport with the phase transition, while $\epsilon_c$ is the gradient energy coefficient. 
From Eq.~\eqref{eq:free_en}, we consider the evolution of the phase field $c$ over time, which yields the well-known Allen-Cahn equation or time-dependent Ginzburg-Landau equation, i.e.,
\begin{equation}
\label{eq:allen_cahn}
- \frac{1}{M_c} \dot{c}
= \frac{\partial \psi_c}{\partial c} - \diver{\left( \frac{\partial \psi_c}{\partial \grad{c}} \right)}
= \frac{\partial f_c}{\partial c} - \epsilon_c^2 \nabla^2 c ,
\end{equation}
where $\nabla^2 ( \bullet ) = \nabla \cdot \grad{( \bullet )}$ is the Laplacian operator and $M_c$ is the mobility parameter. 
Since this study does not consider solute transport or any other chemical effects, we focus on the pure water-ice phase transition such that the free energy density $f_c(\theta,c)$ can be written as,
\begin{equation}
\label{eq:f_c}
f_c = W_c g_c(c) + \mathcal{F}_c(\theta) p_c(c),
\end{equation}
where $g_c(c) = c^2 (1-c)^2$ is the double well potential [Fig. \ref{fig:g_c}] that can be regarded as an energy barrier at the ice-water interface with the height of $W_c$, and $p_c(c) = S^w(c) = c^3 (6c^2 -15c + 10)$ is the interpolation function [Fig. \ref{fig:p_c}], while $\mathcal{F}_c(\theta)$ is the driving force that is a first-order Taylor approximation of the cryo-suction ($s_{\text{cryo}}$) based on the Clausius-Clapeyron relation \citep{boettinger2002phase}:
\begin{equation}
\label{eq:cryo_suction}
s_{\text{cryo}} 
= p_i - p_w 
= - \rho_i L_{\theta} \ln{\frac{\theta}{\theta_m}} 
\approx \mathcal{F}_c(\theta) 
= \rho_i L_{\theta} \left( 1 - \frac{\theta}{\theta_m} \right).
\end{equation}

\begin{figure}[h]
\centering
\includegraphics[height=0.325\textwidth]{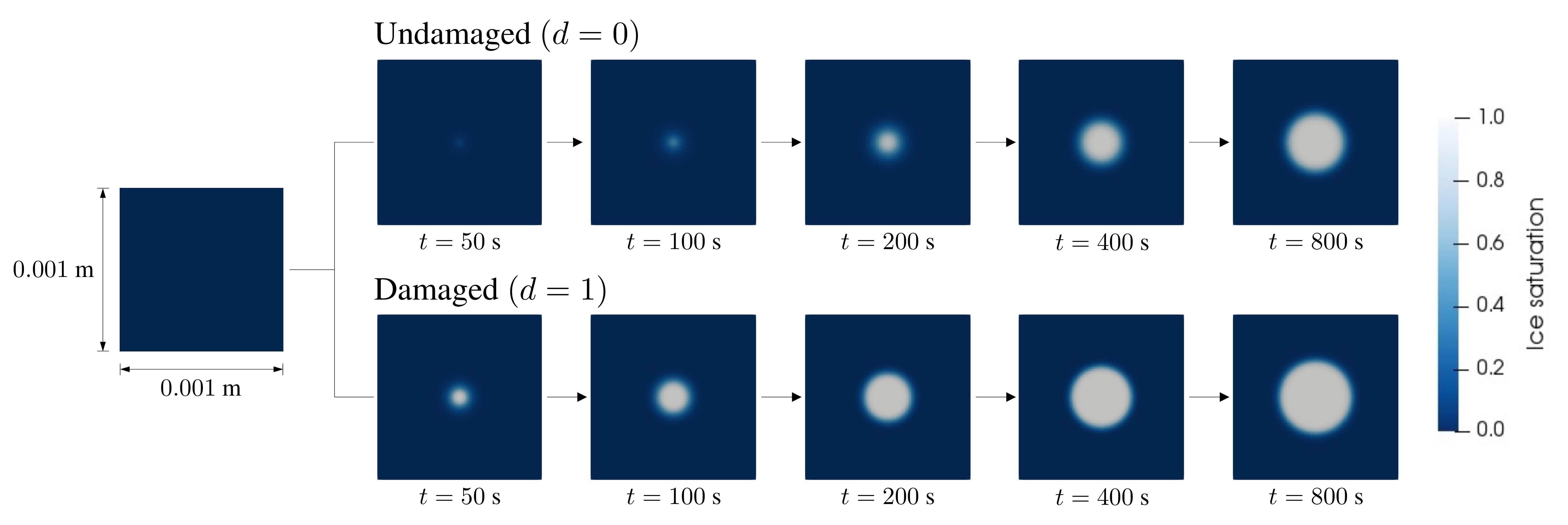}
\caption{Different growth rates of the ice phases when a heat sink of $\hat{r} = -10^9$ W/m$^3$ is placed at a small region at the center with the area of $A_c = 10^{-10}$ m$^2$.}
\label{fig:driving_force_growth}
\end{figure}

\noindent
As pointed out in \citep{warren1995prediction, boettinger2002phase}, since Eq.~\eqref{eq:allen_cahn} captures the evolution of the regularized ice-water interface, numerical parameters $\epsilon_c$, $W_c$, and $M_c$ can be related to the ice-water surface tension $\gamma_{iw}$, the interface thickness $\delta_c$, and the kinetic coefficient $\nu_c$ as,
\begin{equation}
\epsilon_c = \sqrt{6 \gamma_{iw} \delta_c}
\: \: ; \: \:
W_c = \frac{3 \gamma_{iw}}{\delta_c}
\: \: ; \: \:
M_c = \frac{\nu_c \theta_m}{ 6 \rho_i L_{\theta} \delta_c }.
\end{equation}

t
Furthermore, since the existence of segregated ice governs the heave rate of frozen soil \citep{penner1986aspects, michalowski2006frost}, this study considers different rates between homogeneous freezing and ice lens growth. 
Specifically, while employing different volumetric expansion coefficients for the in-pore crystallization and the formation of ice lens [Eq.~\eqref{eq:total_stress}], we replace the driving force $\mathcal{F}_c(\theta)$ with $\mathcal{F}_c^*(\theta,d)$ that contains additional term that describes the intense growth of ice lenses similar to the kinetic equation proposed by Espinosa et al. \citep{espinosa2008phase}, which is often used to model salt crystallization in porous media \citep{koniorczyk2012modelling, derluyn2014deformation, choo2018cracking}: 
\begin{equation}
\label{eq:driving_force}
\mathcal{F}^*_c (\theta, d) = \rho_i L_{\theta} \left( 1 - \frac{\theta}{\theta_m} \right) + [1 - g_d(d)] K^*_c \left( 1 - \frac{\theta}{\theta_m} \right)^{g^*_c},
\end{equation}
where $K^*_c >0$ and $g^*_c >0$ are the kinetic parameters. 
The effect of the additional term in Eq.~\eqref{eq:driving_force} is illustrated in Fig. \ref{fig:driving_force_growth}, where we simulate the water-ice phase transition by placing a heat sink at the center while the kinetic parameters are set to be $K^*_c = 5.0$ GPa and $g^*_c = 1.2$. 
By considering two different cases where the entire 1 mm$^2$ large water-saturated square domain remains intact and is completely damaged, Fig. \ref{fig:driving_force_growth} shows that the modified driving force $\mathcal{F}^*_c$ is capable of capturing different growth rates depending on the damage parameter $d$. 

\subsection{Freezing-induced fracture in microporoelastic medium}
\label{sec:freezing_frac}
\subsubsection{Microporomechanics of the phase-changing porous medium}
\label{sec:poroelasticity}
\begin{figure}[h]
\centering
\includegraphics[height=0.375\textwidth]{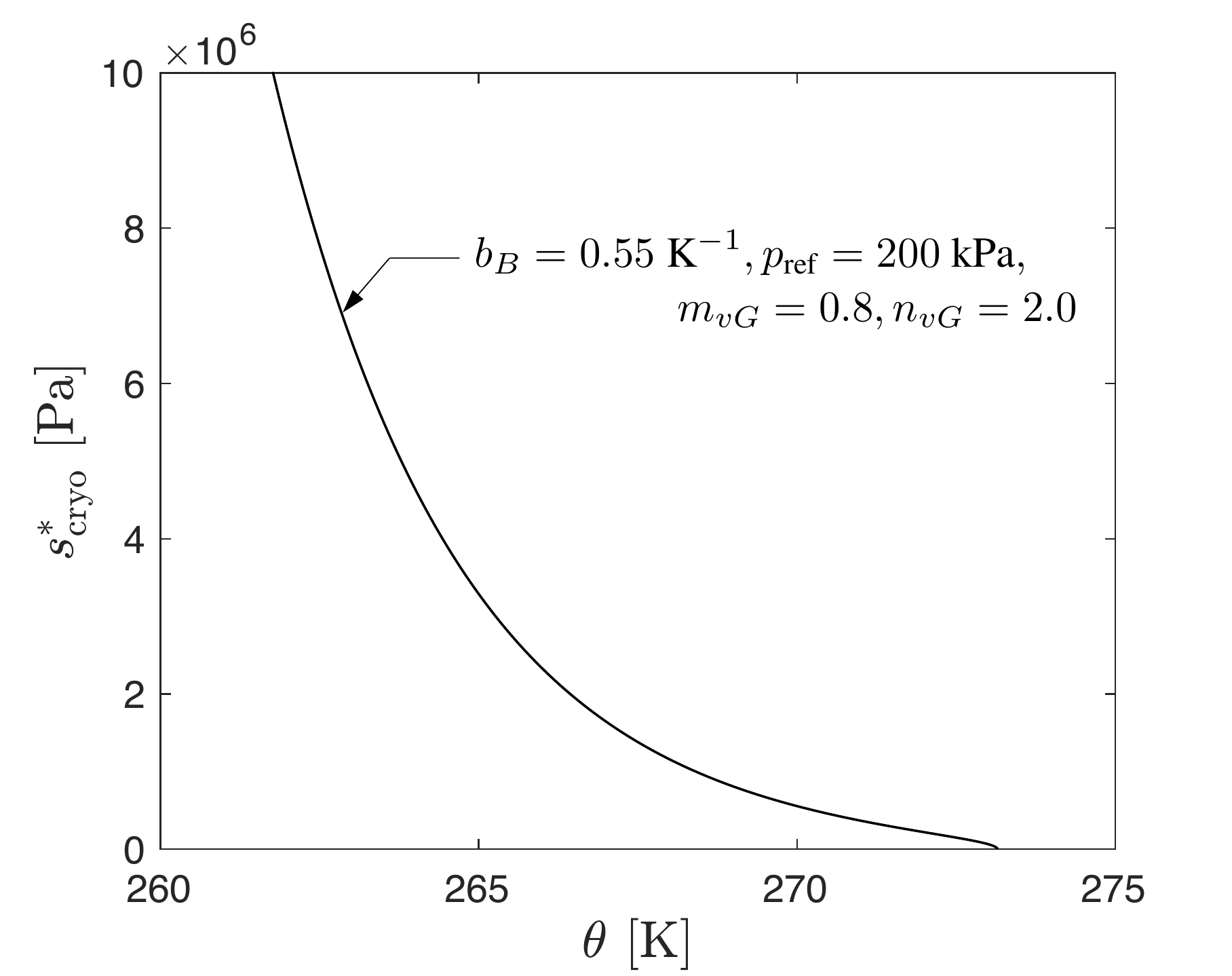}
\caption{Freezing characteristic function [Eq.~\eqref{eq:freezing_retention}] used in this study.}
\label{fig:freez_ret}
\end{figure}

Focusing on the ice lens formation that involves a long period of time up to annual scales \citep{guodong1983mechanism, harris2009permafrost}, this study neglects the inertial effects such that the balance of linear momentum for the three-phase mixture can be written as, 
\begin{equation}
\label{eq:mom_bal_mixture}
\diver{(\bar{\tensor{\sigma}}' - \bar{p}\tensor{I})} + \rho \vec{g} = \vec{0}.
\end{equation}
Based on the observation that geological materials remain brittle at a low temperature \citep{evans1990brittle, lee2002frozen}, we assume that the evolution of the damage parameter $d$ replicates the mechanism of brittle fracture. 
In this case, undamaged effective stress $\tensor{\sigma}'_{\text{int}}$ can be considered linear elastic, while the stress tensor inside the damaged zone should remain $\tensor{\sigma}'_{\text{dam}} = \tensor{0}$ unless the temperature is below $\theta_m$ to form a bulk ice. 
Moreover, since the ice flow with respect to the solid phase is negligible compared to that of water \citep{zhou2013three, na2017computational}, both $\tensor{\sigma}'_{\text{int}}$ and $\tensor{\sigma}'_{\text{dam}}$ can be related to the strain measure $\tensor{\varepsilon} = (\grad{\vec{u}} + \grad{\vec{u}}^{\text{T}})/2$ by approximating $\tilde{\vec{v}}_{i} \approx \vec{0}$. 
Given these considerations, we define the constitutive relations for $\tensor{\sigma}'_{\text{int}}$ and $\tensor{\sigma}'_{\text{dam}}$ as,
\begin{equation}
\label{eq:constitutive_solid}
\tensor{\sigma}'_{\text{int}} 
= K \varepsilon^{\text{vol}} \tensor{I} + 2 G \tensor{\varepsilon}^{\text{dev}}
\: \: ; \: \: 
\tensor{\sigma}'_{\text{dam}} 
= [1 - S^w(c)] (K_i \varepsilon^{\text{vol}} \tensor{I} + 2 G_i \tensor{\varepsilon}^{\text{dev}}),
\end{equation}
where $\varepsilon^{\text{vol}} = \tr{(\tensor{\varepsilon})}$ and $\tensor{\varepsilon}^{\text{dev}} = \tensor{\varepsilon} - (\varepsilon^{\text{vol}}/3)\tensor{I}$, while $K$ and $K_i$ are the bulk moduli; and $G$ and $G_i$ are the shear moduli for the solid skeleton and the ice, respectively. 
Based on this approach, $\tensor{\sigma}'_{\text{dam}}$ can be interpreted as a developed stress due to the ice lens growth, since it not only depends on the fracturing process but also on the state of the fluid phase. 
The net pore pressure $\bar{p}$, on the other hand, is a driver of deformation and fracture due to the formation of ice crystal that exerts significant excess pressure on the premelted water film. 
This pressure is referred to as cryo-suction $s_{\text{cryo}}$ that induces the ice pressure $p_i$ to be far greater than the water pressure $p_w$. 
As shown in Eqs.~\eqref{eq:pore_pressure_partitioning} and \eqref{eq:cryo_suction}, the net pore pressure can be rewritten as $\bar{p} = [1 - S^w(c)]s_{\text{cryo}} - p_w$, while $s_{\text{cryo}}$ can be determined based upon the Clausius-Clapeyron equation. 
In practice, however, the Clausius-Clapeyron equation is typically replaced by an empirical model, such as the exponential \citep{anderson1972predicting} or the van Genuchten \citep{van1980closed} curves, which is considered to be more accurate, evidenced by the experiments \citep{koopmans1966soil, black1989comparison, ma2017soil, bai2018theory}: 
\begin{equation}
\label{eq:empirical}
{S^w}^* = \exp{\left(b_B \langle \theta - \theta_m \rangle_{-}\right)}
\: \: ; \: \:
s_{\text{cryo}}^* = p_{\text{ref}} \left[ \lbrace S^w(c) \rbrace^{-\frac{1}{m_{vG}}} - 1 \right]^{\frac{1}{n_{vG}}},
\end{equation}
where $b_B$, $p_{\text{ref}}$, $m_{vG}$, and $n_{vG}$ are empirical parameters while $ \langle \bullet \rangle_{\pm} = (\bullet \pm |\bullet|)/2$ is the Macaulay bracket. 
Note that we use a superscripted symbol $*$ to indicate that the corresponding variables are empirically determined. 
Yet, these empirical models still yield unrealistic results in some cases. 
For example, the derivative of the exponential model possesses a big jump discontinuity at the freezing temperature $\theta_m$, while $s_{\text{cryo}}^*$ approaches infinity if $S^w(c) \to 0$ if adopting the van Genuchten model. 
Hence, in this study, we combine the two models to obtain the freezing retention curve that bypasses such issues (Fig. \ref{fig:freez_ret}):
\begin{equation}
\label{eq:freezing_retention}
s_{\text{cryo}}^* = p_{\text{ref}} \left\lbrace \left[ \left\lbrace \exp{\left(b_B \langle \theta - \theta_m \rangle_{-}\right)} \right\rbrace \right]^{-\frac{1}{m_{vG}}} - 1 \right\rbrace^{\frac{1}{n_{vG}}}, 
\end{equation}
and we replace $s_{\text{cryo}}$ with $s_{\text{cryo}}^*$ for the net pore pressure such that: $\bar{p} = [1 - S^w(c)]s_{\text{cryo}}^* - p_w$. 
For all the numerical examples presented in Section \ref{sec:example}, we adopt the same values used in \citep{na2017computational, bai2018theory}: $b_B = 0.55$ K$^{-1}$, $p_{\text{ref}} = 200$ kPa, $m_{vG} = 0.8$, and $n_{vG} = 2.0$. 

Recall Section \ref{sec:modeling_appr} that our material of interest is a fluid-saturated phase-changing porous media. 
Thus, this study considers the balance of mass for three phase constituents (i.e., solid, water and ice) as follows:
\begin{align}
\label{eq:mass_bal_s}
& \dot{\rho}^s + \rho^s \diver{\vec{v}} = \dot{m}_s, \\
\label{eq:mass_bal_w}
& \dot{\rho}^w + \rho^w \diver{\vec{v}} + \diver{\rho^w \tilde{\vec{v}}_w} = \dot{m}_w, \\
\label{eq:mass_bal_i}
& \dot{\rho}^i + \rho^i \diver{\vec{v}} + \diver{\rho^i \tilde{\vec{v}}_i} = \dot{m}_i,
\end{align}
where $\dot{m}_s$, $\dot{m}_w$, and $\dot{m}_i$ indicate the mass production rate for each phase constituent \citep{zhou2013three, na2017computational, choo2018cracking}. 
Here, we assume that only the water and ice phase constituents exchange mass among constituents (i.e., $\dot{m}_s = 0$ and $\dot{m}_w = -\dot{m}_i$). 
Hence, summation of Eqs.~\eqref{eq:mass_bal_w} and \eqref{eq:mass_bal_i} yields:
\begin{equation}
\label{eq:sum_bal_w_i}
\dot{\phi} \left\lbrace S^w(c) \rho_w + [1-S^w(c)] \rho_i \right\rbrace + \phi \dot{S}^w(c) (\rho_w - \rho_i)
+
\phi \left\lbrace S^w(c) \rho_w + [1 - S^w(c)] \rho_i \right\rbrace \diver{\vec{v}}
+
\diver{\rho^w \tilde{\vec{v}}_w} = 0,
\end{equation}
since $\tilde{\vec{v}}_i \approx \vec{0}$, while Eq.~\eqref{eq:mass_bal_s} can be rewritten as,
\begin{equation}
\label{eq:mass_bal_s2}
\dot{\phi} = (1 - \phi) \diver{\vec{v}}.
\end{equation}
Substituting Eq.~\eqref{eq:mass_bal_s2} into \eqref{eq:sum_bal_w_i} yields the mass balance equation for the three-phase mixture:
\begin{equation}
\label{eq:mass_bal_mixture}
\phi \dot{S}^w(c) (\rho_w - \rho_i)
+
\left\lbrace S^w(c) \rho_w + [1 - S^w(c)] \rho_i \right\rbrace \diver{\vec{v}}
+
\diver{\rho^w \tilde{\vec{v}}_w} = 0.
\end{equation}
In this study, we focus on the case where the water flow inside both the porous matrix and the fracture obeys the generalized Darcy's law while considering the pore blockage due to the water-ice phase transition \citep{luckner1989consistent, seyfried1997use, demand2019influences}. 
In other words, we adopt the following constitutive relation between $\tilde{\vec{v}}_w$ and $p_w$:
\begin{equation}
\label{eq:darcy}
\vec{w}_w = - \frac{k_r \tensor{k}}{\mu_w} ( \grad{p}_w - \rho_w \vec{g}),
\end{equation}
where $\vec{w}_{w} = \phi \tilde{\vec{v}}_w$ is Darcy's velocity, $\tensor{k}$ is the permeability tensor, $\mu_w$ is the water viscosity, and $k_r$ is the saturation dependent relative permeability:
\begin{equation}
\label{eq:rel_perm}
k_r = S^w(c)^{1/2} \left\lbrace 1 - \left[ 1 - S^w(c)^{1/m_{vG}} \right]^{m_{vG}} \right\rbrace^2.
\end{equation}

\subsubsection{Damage evolution}
\label{sec:dam_evol}
Following \citep{suh2021asynchronous}, this study interprets cracking as the fracture of the solid skeleton. 
In other words, we define the crack driving force $\mathcal{F}_d \ge 0$ as,
\begin{equation}
\label{eq:crack_driv_force}
\mathcal{F}_d = - \frac{\partial g_d(d)}{\partial d} \psi'_{\text{int}}
\: \: ; \: \:
\psi'_{\text{int}} = \frac{1}{2}K (\varepsilon^{\text{vol}})^2 + G (\tensor{\varepsilon}^{\text{dev}} : \tensor{\varepsilon}^{\text{dev}}),
\end{equation}
such that the damage evolution equation can be obtained from the balance between the crack driving force $\mathcal{F}_d$ and the crack resistance $\mathcal{R}_d$ \citep{dittmann2019variational, dittmann2020phase, suh2021asynchronous}:
\begin{equation}
\label{eq:phase_field_evol1}
\mathcal{R}_d - \mathcal{F}_d = \frac{\partial g_d(d)}{\partial d} \psi'_{\text{int}} + \frac{\mathcal{G}_d}{l_d} (d - l_d^2 \nabla^2 d) = 0.
\end{equation}
Recall Section \ref{sec:multi_phase_field} that our choice of degradation function $g_d(d)$ reduces the thermodynamic restriction into $\dot{d} \ge 0$ \citep{bryant2018mixed, suh2021asynchronous}, which requires additional treatment to ensure monotonic crack growth.
In this study, we adopt the same treatment used in \citep{miehe2015minimization, bryant2018mixed}. 
By considering the homogeneity $\grad{d} = \vec{0}$, Eq.~\eqref{eq:phase_field_evol1} yields the following expression:
\begin{equation}
\label{eq:d_dot}
\dot{d} = \frac{2}{(1 + 2\mathcal{H})^2} \dot{\mathcal{H}} \ge 0
\: \: ; \: \:
\mathcal{H} = \frac{\psi'_{\text{int}}}{\mathcal{G}_d/l_d},
\end{equation}
implying that non-negative $\dot{d}$ is guaranteed if $\dot{\mathcal{H}} \ge 0$. 
As a simple remedy, we replace $\mathcal{H}$ with $\mathcal{H}^*$ which is defined as the pseudo-temporal maximum of normalized strain energy, while considering a critical value $\mathcal{H}_{\text{crit}}$ that restricts the crack to initiate above a threshold strain energy \citep{miehe2015phase, bryant2018mixed, suh2019open, bryant2021phase}:
\begin{equation}
\label{eq:history}
\mathcal{H}^* = \max_{\tau \in [0,t]} \left\langle \mathcal{H} - \mathcal{H}_{\text{crit}} \right\rangle_{+}, 
\end{equation}
such that Eq.~\eqref{eq:phase_field_evol1} accordingly becomes:
\begin{equation}
\label{eq:phase_field_evol2}
\frac{\partial g_d(d)}{\partial d} \mathcal{H}^* + (d - l_d^2 \nabla^2 d) = 0.
\end{equation}

In order to model the fracture flow in a fluid-infiltrating porous media, we adopt the permeability enhancement approach that approximates the water flow inside the fracture as the flow between two parallel plates \citep{miehe2016phase, mauthe2017hydraulic, wang2017unified, suh2021immersed}:
\begin{equation}
\label{eq:aniso_perm}
\tensor{k} 
= \tensor{k}_{\text{mat}} + \tensor{k}_d 
= k_{\text{mat}} \tensor{I} + d^2 k_d ( \tensor{I} - \vec{n}_d \otimes \vec{n}_d),
\end{equation}
where $k_{\text{mat}}$ is the effective permeability of the undamaged matrix, $\vec{n}_d = \grad{d} / \| \grad{d}\|$ is the unit normal of crack surface, and $k_d = w_d^2/12$ describes the permeability enhancement due to the crack opening which depends on the hydraulic aperture $w_d$ based on the cubic law.  
However, freezing-induced fracture involves different situations where the pore ice crystal growth drives fracture but at the same time blocks the pore that may hinder the water flow therein. 
Hence, we adopt the approach used in \citep{choo2018cracking} which assumes a linear relationship between the hydraulic aperture $w_d$ and the water saturation $S^w(c)$:
\begin{equation}
\label{eq:aperture}
w_d = S^w(c) l_{\perp} ( \vec{n}_d \cdot \tensor{\varepsilon} \cdot \vec{n}_d ),
\end{equation}
where $l_{\perp}$ is the characteristic length of a line element perpendicular to the fracture which is often assumed to be equivalent to the mesh size \citep{miehe2016phase, wilson2016phase}. 
Furthermore, by assuming that the crack opening leads to complete fragmentation of the solid matrix, we adopt the following relation for the porosity \citep{heider2020phase,suh2021asynchronous}: 
\begin{equation}
\label{eq:dam_dep_poro}
\phi = 1 - g_d(d)(1 - \phi_0)(1 - \diver{\vec{u}}),
\end{equation}
such that the porosity approaches 1 if the solid skeleton is completely damaged.

\section{Finite element implementation}
\label{sec:implementation}
This section presents a finite element discretization of the set of governing equations described in Section \ref{sec:microporomechanics}, and the solution strategy for the resulting discrete system. 
We first formulate the weak form of the field equations by following the standard weighted residual procedure. 
In specific, we adopt the Taylor-Hood element for the displacement and pore water pressure fields, while employing linear interpolation functions for all other variables in order to remove spurious oscillations. 
We then describe the operator split solution scheme that separately updates $\lbrace \theta, c \rbrace$ and $\lbrace \vec{u}, p_w \rbrace$, while the damage parameter $d$ is updated in a staggered manner for numerical robustness.

\subsection{Galerkin form}
\label{sec:galerkin_form}
Let domain $\mathcal{B}$ possesses boundary surface $\partial \mathcal{B}$ composed of Dirichlet boundaries (displacement $\partial \mathcal{B}_u$, pore water pressure $\partial \mathcal{B}_p$, and temperature $\partial \mathcal{B}_{\theta}$) and Neumann boundaries (traction $\partial \mathcal{B}_t$, water mass flux $\partial \mathcal{B}_w$, and heat flux $\partial \mathcal{B}_{q}$) that satisfies:
\begin{equation}
\label{eq:boundary}
\partial \mathcal{B} 
= \overline{\partial \mathcal{B}_u \cup \partial \mathcal{B}_t} 
= \overline{\partial \mathcal{B}_p \cup \partial \mathcal{B}_w}
= \overline{\partial \mathcal{B}_{\theta} \cup \partial \mathcal{B}_{q}}
\: \: ; \: \:
\emptyset 
= \partial \mathcal{B}_u \cap \partial \mathcal{B}_t
= \partial \mathcal{B}_p \cap \partial \mathcal{B}_w
= \partial \mathcal{B}_{\theta} \cap \partial \mathcal{B}_{q}. 
\end{equation}
Then, the prescribed boundary conditions can be specified as,
\begin{align}
\label{eq:BCs}
\begin{dcases}
\vec{u} = \hat{\vec{u}} & \text{ on } \partial \mathcal{B}_u, \\
p_w = \hat{p}_w & \text{ on } \partial \mathcal{B}_p, \\
\theta = \hat{\theta} & \text{ on } \partial \mathcal{B}_{\theta},
\end{dcases}
\: \: ; \: \:
\begin{dcases}
\tensor{\sigma} \cdot \vec{n} = \hat{\vec{t}} & \text{ on } \partial \mathcal{B}_t, \\
- \vec{w}_w \cdot \vec{n} = \hat{w}_w & \text{ on } \partial \mathcal{B}_w, \\
- \vec{q} \cdot \vec{n} = \hat{q} & \text{ on } \partial \mathcal{B}_{q},
\end{dcases}
\end{align}
where $\vec{n}$ is the outward-oriented unit normal on the boundary surface $\partial \mathcal{B}$. 
Meanwhile, the following boundary conditions on $\partial \mathcal{B}$ are prescribed for the phase fields $c$ and $d$:
\begin{equation}
\label{eq:BC_c_and_d}
\grad{c} \cdot \vec{n} = 0
\: \: ; \: \:
\grad{d} \cdot \vec{n} = 0.
\end{equation}
For model closure, the initial conditions for the primary unknowns $\lbrace \vec{u}, p_w, \theta, c, d \rbrace$ are imposed as: 
\begin{equation}
\label{eq:IC}
\vec{u} = \vec{u}_0 
\: \: ; \: \:
p_w = p_{w0}
\: \: ; \: \:
\theta = \theta_0
\: \: ; \: \:
c = c_0
\: \: ; \: \:
d = d_0,
\end{equation}
at time $t = 0$. 
We also define the trial spaces $V_u$, $V_p$, $V_{\theta}$, $V_c$, and $V_d$ for the solution variables as,
\begin{equation}
\label{eq:trial_space}
\begin{aligned}
&V_u = \left\lbrace \vec{u} : \mathcal{B} \to \mathbb{R}^3 \: | \: \vec{u} \in [ H^1 ( \mathcal{B} ) ]^3, \: \left. \vec{u} \right|_{\partial \mathcal{B}_u} = \hat{\vec{u}} \right\rbrace, \\
&V_p = \left\lbrace p_w : \mathcal{B} \to \mathbb{R} \: | \: p_w \in H^1 ( \mathcal{B} )  , \: \left. p_w \right|_{\partial \mathcal{B}_p} = \hat{p}_w \right\rbrace, \\
&V_{\theta} = \left\lbrace \theta : \mathcal{B} \to \mathbb{R} \: | \: \theta \in H^1 ( \mathcal{B} )  , \: \left. \theta \right|_{\partial \mathcal{B}_{\theta}} = \hat{\theta} \right\rbrace, \\
&V_c = \left\lbrace c : \mathcal{B} \to \mathbb{R} \: | \: c \in H^1 ( \mathcal{B} ) \right\rbrace, \\
&V_d = \left\lbrace d : \mathcal{B} \to \mathbb{R} \: | \: d \in H^1 ( \mathcal{B} ) \right\rbrace,
\end{aligned}
\end{equation}
which is complimented by the admissible spaces:
\begin{equation}
\label{eq:ad_space}
\begin{aligned}
&V_{\eta} = \left\lbrace \vec{\eta} : \mathcal{B} \to \mathbb{R}^3 \: | \: \vec{\eta} \in [ H^1 ( \mathcal{B} ) ]^3 , \: \left. \vec{\eta} \right|_{\partial \mathcal{B}_u} = \vec{0} \right\rbrace, \\
&V_{\xi} = \left\lbrace \xi : \mathcal{B} \to \mathbb{R} \: | \: \xi \in H^1 ( \mathcal{B} )  , \: \left. \xi \right|_{\partial \mathcal{B}_p} = 0 \right\rbrace, \\
&V_{\zeta} = \left\lbrace \zeta : \mathcal{B} \to \mathbb{R} \: | \: \zeta \in H^1 ( \mathcal{B} )  , \: \left. \zeta \right|_{\partial \mathcal{B}_{\theta}} = 0 \right\rbrace, \\
&V_{\gamma} = \left\lbrace \gamma : \mathcal{B} \to \mathbb{R} \: | \: \gamma \in H^1 ( \mathcal{B} ) \right\rbrace, \\
&V_{\omega} = \left\lbrace \omega : \mathcal{B} \to \mathbb{R} \: | \: \omega \in H^1 ( \mathcal{B} ) \right\rbrace,
\end{aligned}
\end{equation}
where $H^1$ indicates the Sobolev space of order 1. 
Specifically, we employ the inf-sup stable Taylor-Hood finite element for the displacement and pore water pressure fields, while all other field variables are discretized with linear functions that eliminates spurious oscillations \citep{borja1998elastoplastic, zienkiewicz1999computational}. 
By applying the standard weighted residual procedure, the weak statements for Eqs.~\eqref{eq:en_bal2}, \eqref{eq:allen_cahn}, \eqref{eq:mom_bal_mixture}, \eqref{eq:mass_bal_mixture}, and \eqref{eq:phase_field_evol2} are to: find $\lbrace \vec{u}, p_w, \theta, c, d \rbrace \in V_u \times V_p \times V_{\theta} \times V_c \times V_d$ such that for all $\lbrace \vec{\eta}, \xi, \zeta, \gamma, \omega \rbrace \in V_{\eta} \times V_{\xi} \times V_{\zeta} \times V_{\gamma} \times V_{\omega}$,
\begin{equation}
\label{eq:weak_form}
G_u = G_p = G_{\theta} = G_c = G_d = 0,
\end{equation}
where:
\begin{empheq}{align}
\label{eq:weak_u}
G_u 
&= 
\int_{\mathcal{B}} \grad{\vec{\eta}} : \tensor{\sigma} \: dV 
- \int_{\mathcal{B}} \vec{\eta} \cdot \rho \vec{g} \: dV
- \int_{\partial \mathcal{B}_t} \vec{\eta} \cdot \hat{\vec{t}} \: d\Gamma = 0, \\ 
\label{eq:weak_p}
\begin{split}
G_p 
&= 
\int_{\mathcal{B}} \xi \left[ \phi \dot{S}^w(c) (\rho_w - \rho_i) \right] \: dV
+ \int_{\mathcal{B}} \xi \left\lbrace S^w(c) \rho_w + [1 - S^w(c)] \rho_i \right\rbrace \diver{\vec{v}} \: dV \\
&
- \int_{\mathcal{B}} \grad{\xi} \cdot (\rho_w \vec{w}_w) \: dV
- \int_{\partial \mathcal{B}_w} \xi (\rho_w \hat{\vec{w}}_w) \: d\Gamma = 0,
\end{split}\\
\label{eq:weak_theta}
\begin{split}
G_{\theta} 
&= 
\int_{\mathcal{B}} \zeta ( \rho^s c_s + \rho^w c_w + \rho^i c_i ) \dot{\theta} \: dV
+ \int_{\mathcal{B}} \zeta \left\lbrace \phi \left[ (\rho_w c_w - \rho_i c_i)(\theta - \theta_m) + \rho_i L_{\theta} \right] \dot{S}^w(c) \right\rbrace \: dV \\
&
- \int_{\mathcal{B}} \grad{\zeta} \cdot \vec{q} \: dV
- \int_{\mathcal{B}} \zeta \hat{r} \: dV
- \int_{\partial \mathcal{B}_q} \zeta \hat{q} \: d\Gamma = 0,
\end{split}\\
\label{eq:weak_c}
G_c 
&= \int_{\mathcal{B}} \gamma \frac{1}{M_c} \dot{c} \: dV
+ \int_{\mathcal{B}} \gamma \frac{\partial f_c}{\partial c} \: dV
+ \int_{\mathcal{B}} \grad{\gamma} \cdot (\epsilon_c^2 \grad{c}) \: dV = 0, \\
\label{eq:weak_d}
G_d 
&= 
\int_{\mathcal{B}} \omega \frac{\partial g_d(d)}{\partial d} \mathcal{H}^* \: dV
+ \int_{\mathcal{B}} \omega d \: dV
+ \int_{\mathcal{B}} \grad{\omega} \cdot (l_d^2 \grad{d}) \: dV = 0.
\end{empheq}

\subsection{Operator-split solution strategy}
\label{sec:operator_split}
Although one may consider different strategies to solve the coupled system of equations [Eqs.~\eqref{eq:weak_u}-\eqref{eq:weak_d}], the solution strategy adopted in this study combines the staggered scheme \citep{miehe2010phase} and the isothermal operator splitting scheme \citep{simo1992associative, nguyen1995coupled}. 
Specifically, we first update the damage field $d$ via linear solver while the variables $\lbrace \vec{u}, p_w, \theta, c \rbrace$ are held fixed. 
We then apply the isothermal splitting solution scheme that iteratively solves the thermally-induced phase transition problem to advance $\lbrace \theta, c \rbrace$, followed by a linear solver that updates $\lbrace \vec{u}, p_w \rbrace$ by solving an isothermal poromechanics problem \citep{suh2021asynchronous}, i.e.,
\begin{equation}
\label{eq:solution_strategies}
\begin{bmatrix}
\vec{u}_{\text{n}} \\
p_{w,\text{n}} \\
\theta_{\text{n}} \\
c_{\text{n}} \\
d_{\text{n}}
\end{bmatrix} \underbrace{
\xrightarrow[\delta \vec{u} = \vec{0}, \: \delta p_w = 0, \: \delta \theta = 0, \: \delta c = 0]{G_d = 0}
\begin{bmatrix}
\vec{u}_{\text{n}} \\
p_{w,\text{n}} \\
\theta_{\text{n}} \\
c_{\text{n}} \\
d_{\text{n}+1}
\end{bmatrix}}_{\text{Linear solver}}
\overbrace{
\xrightarrow[\delta \vec{u} = \vec{0}, \: \delta p_w = 0, \: \delta d = 0]{G_{\theta} = G_c = 0}
\begin{bmatrix}
\vec{u}_{\text{n}} \\
p_{w,\text{n}} \\
\theta_{\text{n}+1} \\
c_{\text{n}+1} \\
d_{\text{n}+1}
\end{bmatrix}}^{\text{Iterative solver}}
\underbrace{
\xrightarrow[\delta \theta = 0, \: \delta c = 0, \: \delta d = 0]{G_u = G_p = 0}
\begin{bmatrix}
\vec{u}_{\text{n}+1} \\
p_{w,\text{n}+1} \\
\theta_{\text{n}+1} \\
c_{\text{n}+1} \\
d_{\text{n}+1}
\end{bmatrix}}_{\text{Linear solver}},
\end{equation}
where we adopt an implicit backward Euler time integration scheme. 
The implementation of the model including the finite element discretization and the solution scheme relies on the finite element package \verb|FEniCS| \citep{logg2010dolfin, logg2012automated, alnaes2015fenics} with \verb|PETSc| scientific computational toolkit \citep{abhyankar2018petsc}.

\section{Numerical examples}
\label{sec:example}
This section presents three sets of numerical examples to verify (Section \ref{sec:verification}), validate (Section \ref{sec:validation}), and showcase (Section \ref{sec:ice_lens}) the capacity of the proposed model. 
Since the evolution of two phase fields $c$ and $d$ requires a fine mesh to capture their sharp gradients, we limit our attention to one- or two-dimensional simulations while considering the diffusion coefficient $\epsilon_c$ as an individual input parameter independent to the interface thickness $\delta_c$ which may additionally reduce the computational cost \citep{sweidan2020unified, sweidan2021experimental}. 
We first present two examples that simulate the latent heat effect and 1d consolidation to verify the implementation of our proposed model. 
As a validation exercise, we perform numerical experiments that replicate the physical experiments conducted by Feng et al. \citep{feng2015unidirectional}, which studies the homogeneous freezing of a phase change material (PCM) embedded in metal foams. 
Then, our final example showcases the performance of the computational model for simulating the ice lens formation and the thermo-hydro-mechanical processes in geomaterials undergoing freeze-thaw cycle.

\subsection{Verification exercises: latent heat effect and 1d consolidation}
\label{sec:verification}
\begin{figure}[h]
\centering
\subfigure[]{\label{fig:lackner_sch}\includegraphics[height=0.25\textwidth]{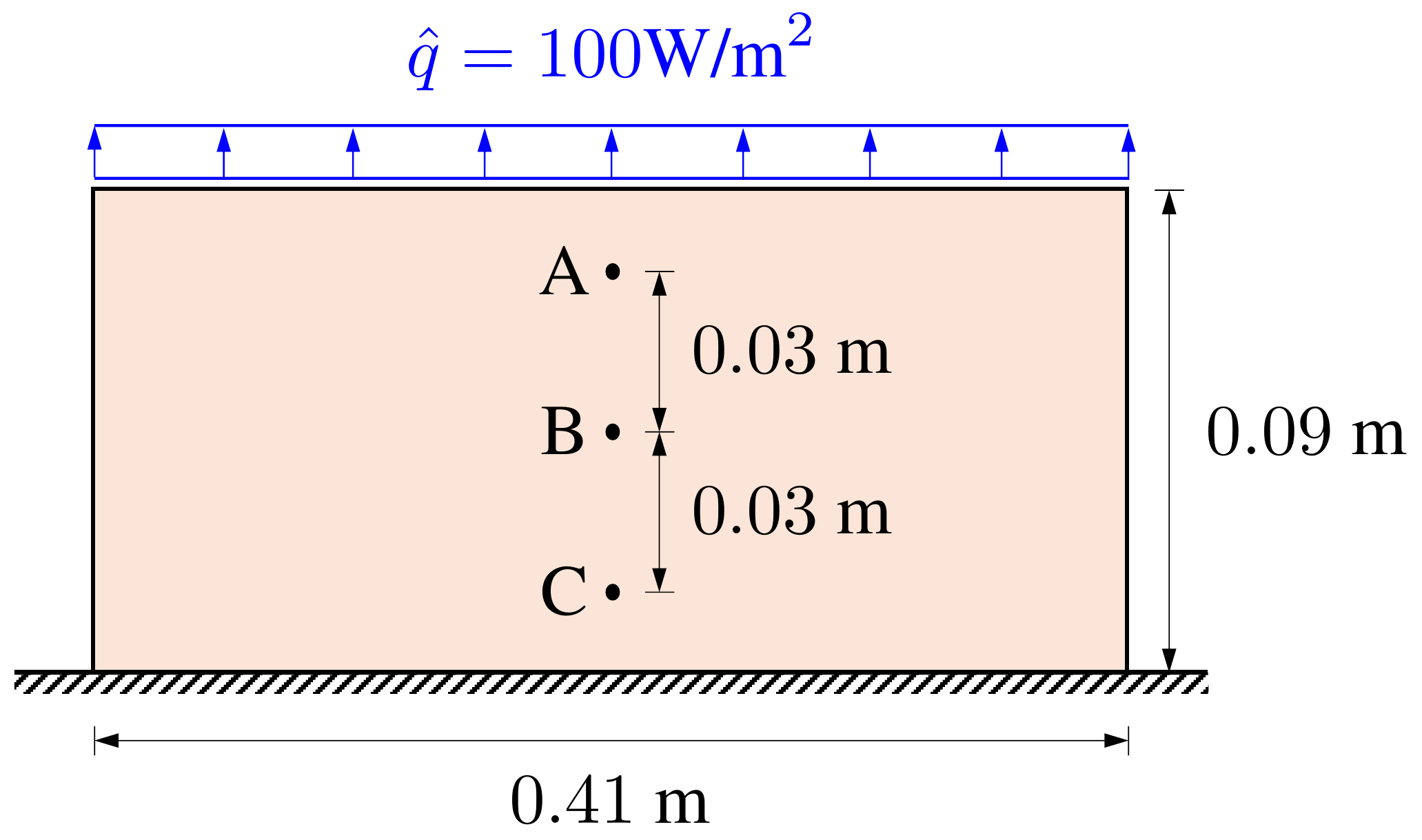}}
\hspace{0.01\textwidth}
\subfigure[]{\label{fig:lackner_temp}\includegraphics[height=0.375\textwidth]{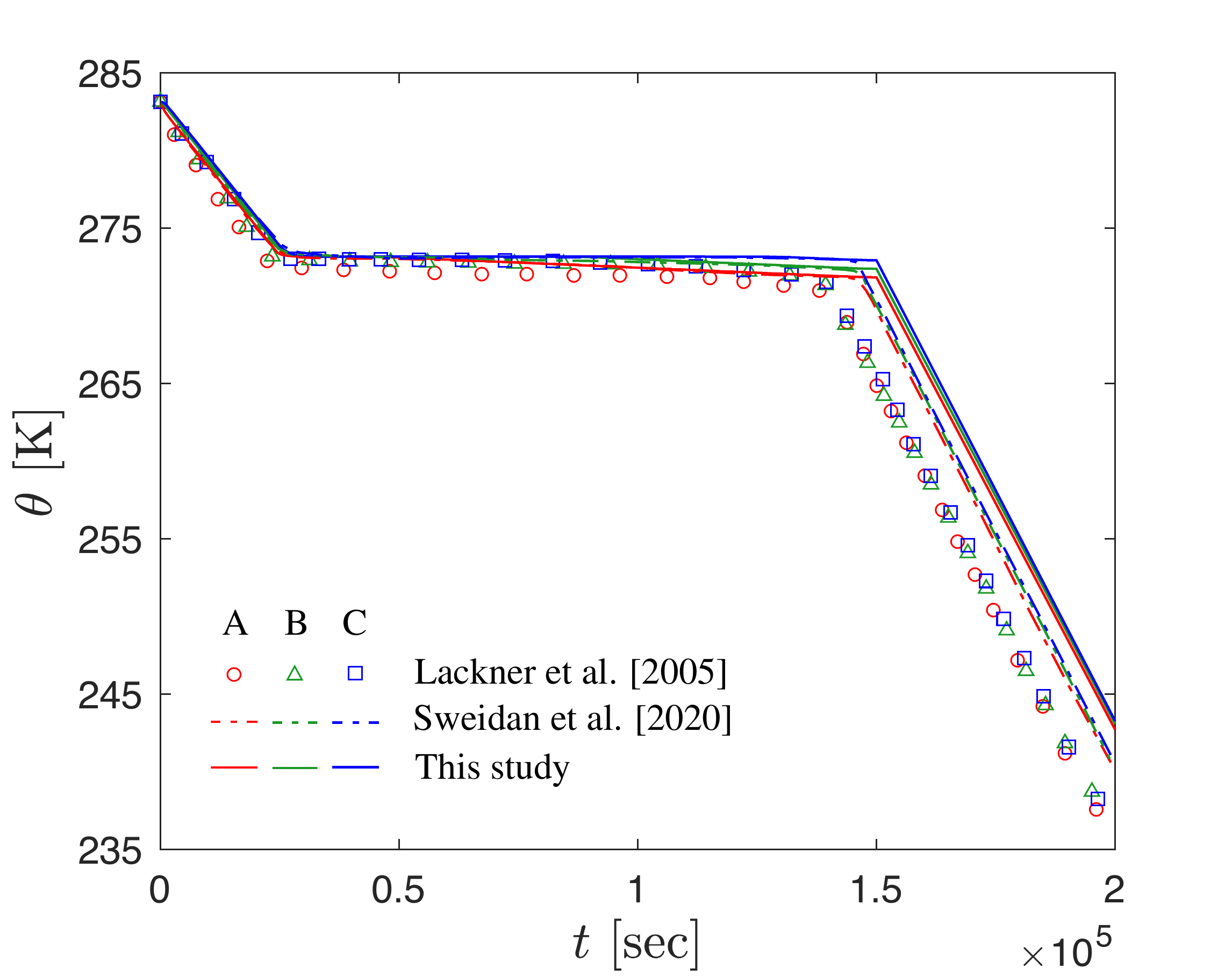}}
\caption{(a) Schematic of geometry and boundary conditions for the 1d freezing example; (b) Temperature evolution at points A, B, and C.}
\label{fig:lackner}
\end{figure}

Our first example simulates one-dimensional freezing of water-saturated porous media to investigate the phase transition of the fluid phase $\alpha = \lbrace w, i \rbrace$ and the involved latent heat effect. 
By comparing the results against the models presented by Lackner et al. \citep{lackner2005artificial} and Sweidan et al. \citep{sweidan2020unified}, this example serves as a verification exercise that ensures the robust implementation of the heat transfer model involving phase transition [i.e., Eqs.~\eqref{eq:weak_theta} and \eqref{eq:weak_c}]. 
Hence, this example considers a rigid solid matrix while neglecting the fluid flow, following \citep{lackner2005artificial}. 
As illustrated in Fig. \ref{fig:lackner_sch}, the problem domain is a fully saturated rectangular specimen with a height of 0.09 m and a width of 0.41 m. 
While the initial temperature of the entire specimen is set to be $\theta_0 = 283.15$ K, the specimen is subjected to freezing with a constant heat flux of $\hat{q} = 100$ W/m$^2$ on the top surface, whereas all other boundaries are thermally insulated. 
Here, we choose the same material properties used in \citep{lackner2005artificial} and \citep{sweidan2020unified} as follows: $\phi_0 = 0.42$, $\rho_s = 2650$ kg/m$^3$, $\rho_w = 1000$ kg/m$^3$, $\rho_i = 913$ kg/m$^3$, $c_s = 740$ J/kg/K, $c_w = 4200$ J/kg/K, $c_i = 1900$ J/kg/K, $\kappa_s = 7.694$ W/m/K, $\kappa_w = 0.611$ W/m/K, and $\kappa_i = 2.222$ W/m/K. 
In addition, we set $\nu_c = 0.001$ m/s, $\gamma_c = 0.03$ J/m$^2$, $\delta_c = 0.005$ m, and $\epsilon_c = 1.25$ (J/m)$^{1/2}$ for the Allen-Cahn phase field model, while we use the structured mesh with element size of $h_e = 0.6$ mm and choose the time step size of $\Delta t = 100$ sec.

As shown in Figure \ref{fig:lackner_temp}, measured temperatures at points A, B, and C during the simulation first linearly decrease due to the applied heat flux $\hat{q}$ until they reach the freezing temperature of $\theta_m = 273.15$ K. 
As soon as the phase transition starts, the freezing front propagates through the specimen while the release of the energy associated with the phase transition prevents the temperature decrease (i.e., latent heat effect). 
Once the phase change is complete, the temperature linearly decreases over time again since the heat transfer process is no longer affected by the latent heat. 
More importantly, a good agreement with the results reported in \citep{lackner2005artificial, sweidan2020unified} verifies that our proposed model is capable of capturing the thermal behavior of the phase-changing porous media. 

\begin{figure}[h]
\centering
\subfigure[]{\label{fig:terzaghi_sch}\includegraphics[height=0.42\textwidth]{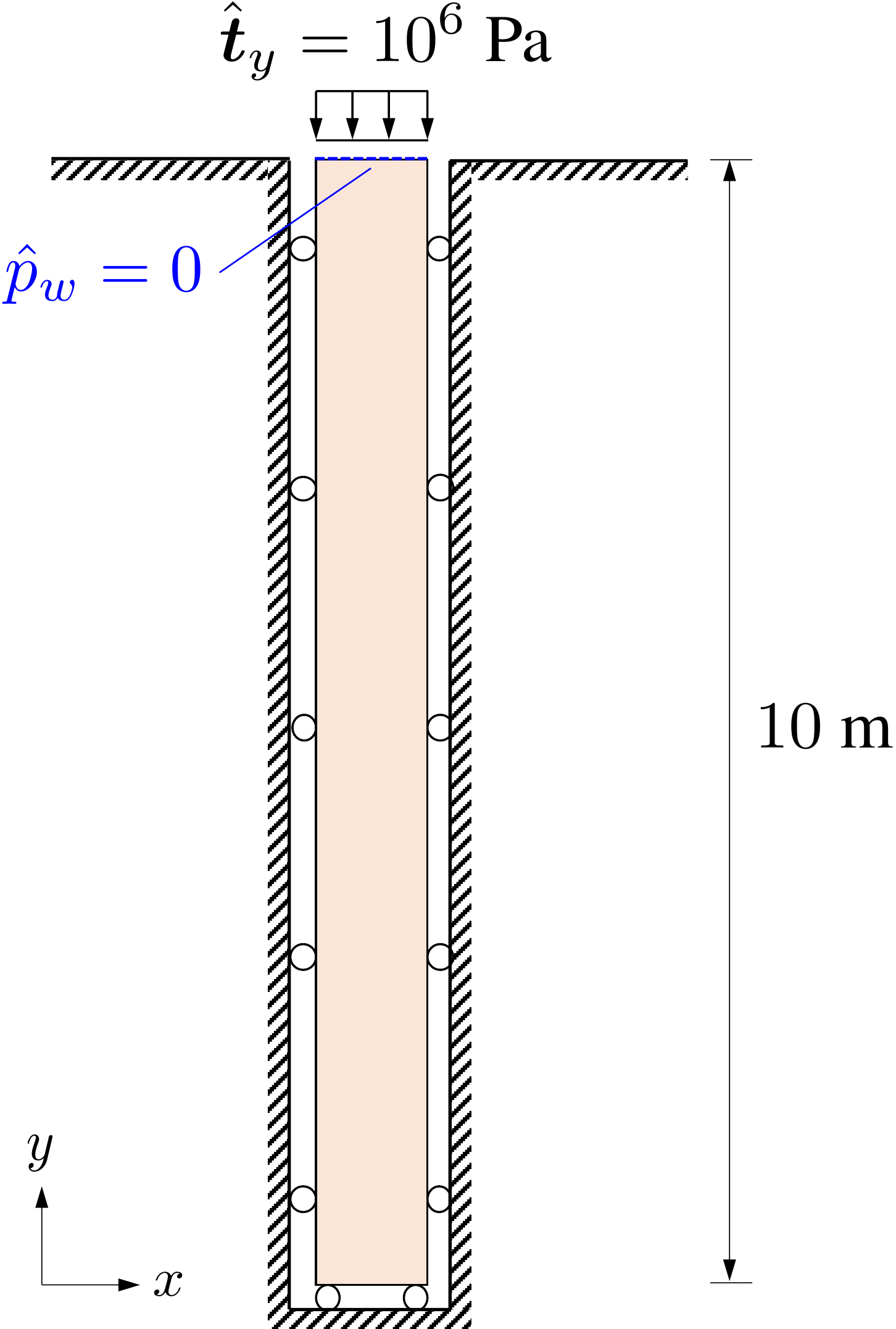}}
\hspace{0.01\textwidth}
\subfigure[]{\label{fig:terzaghi_pw}\includegraphics[height=0.375\textwidth]{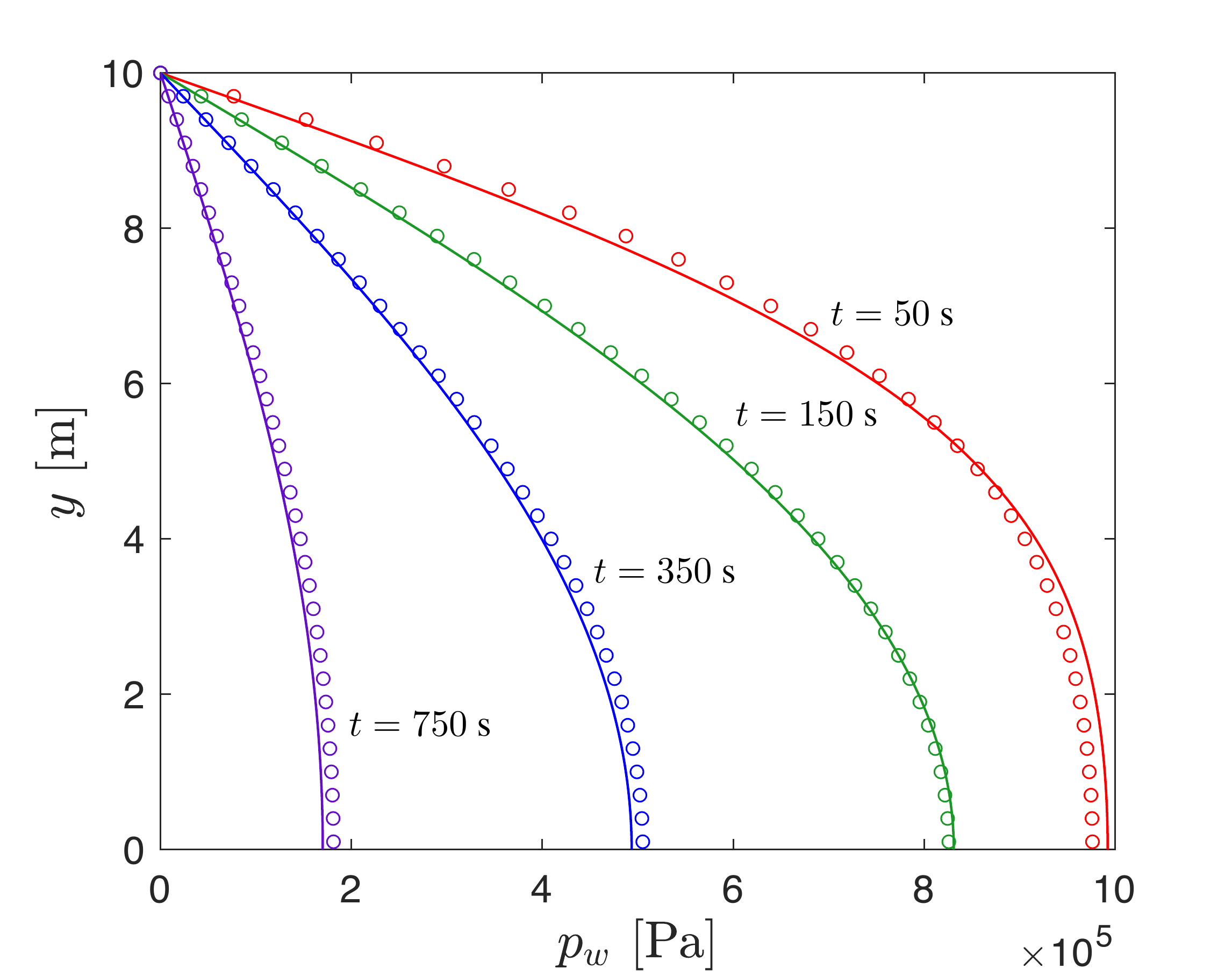}}
\caption{(a) Schematic of geometry and boundary conditions for Terzaghi's problem; (b) Time-dependent pore water pressures along the height of the specimen.}
\label{fig:terzaghi}
\end{figure}

For the second verification exercise, we choose the classical Terzaghi's 1d consolidation problem since it possesses an analytical solution \citep{terzaghi1996soil}, which can directly be compared with the results obtained via poromechanics model [Eqs.~\eqref{eq:weak_u} and \eqref{eq:weak_p}]. 
Our problem domain shown in Fig. \ref{fig:terzaghi_sch} consists of a 10 m high water-saturated linear elastic soil mass. 
While a 1 MPa compressive load $\vec{t}_y$ is imposed on the top surface, we replicate the single-drained condition by prescribing zero pore water pressure at the top ($\hat{p}_w = 0$) and a no-slip condition at the bottom. 
By assuming that the temperature of the soil column remains constant during the simulation ($\theta = 293.15$ K), we only focus on its hydro-mechanically coupled response while the material parameters are chosen as follows: $\phi_0 = 0.4$, $\rho_s = 2650$ kg/m$^3$, $\rho_w = 1000$ kg/m$^3$, $K = 66.67$ MPa, $G = 40$ MPa, $k_{\text{mat}} = 10^{-12}$ m$^2$, and $\mu_w = 10^{-3}$ Pa$\cdot$s. 
Here, we choose $h_e = 0.1$ m and $\Delta t = 20$ sec.

Fig. \ref{fig:terzaghi_pw} illustrates the pore water pressure profile during the simulation at $t = 50$, 150, 350, and 750 s. 
The results show that the applied mechanical load $\vec{t}_y$ builds up the pore water pressure, affecting the pore water to migrate towards the top surface, which leads to the dissipation of the excess pressure over time (i.e., consolidation). 
By comparing the simulation results (circular symbols) to the analytical solution (solid curves), Fig. \ref{fig:terzaghi_pw} verifies the reliability of our model to capture the hydro-mechanically coupled responses.

\begin{figure}[h]
\centering
\subfigure[]{\label{fig:feng_sch}\includegraphics[height=0.46\textwidth]{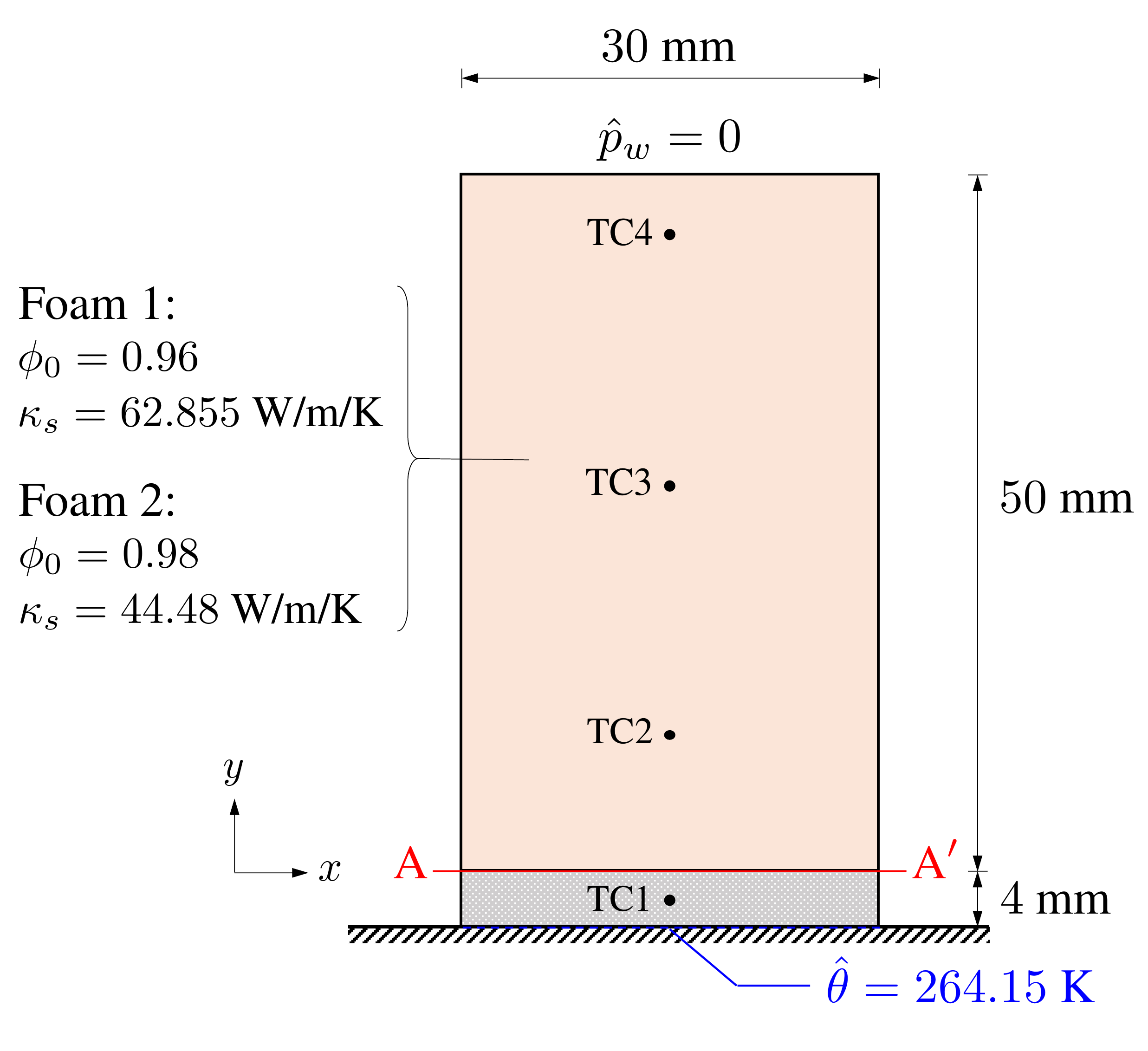}}
\hspace{0.01\textwidth}
\subfigure[]{\label{fig:feng_applied_T}\includegraphics[height=0.375\textwidth]{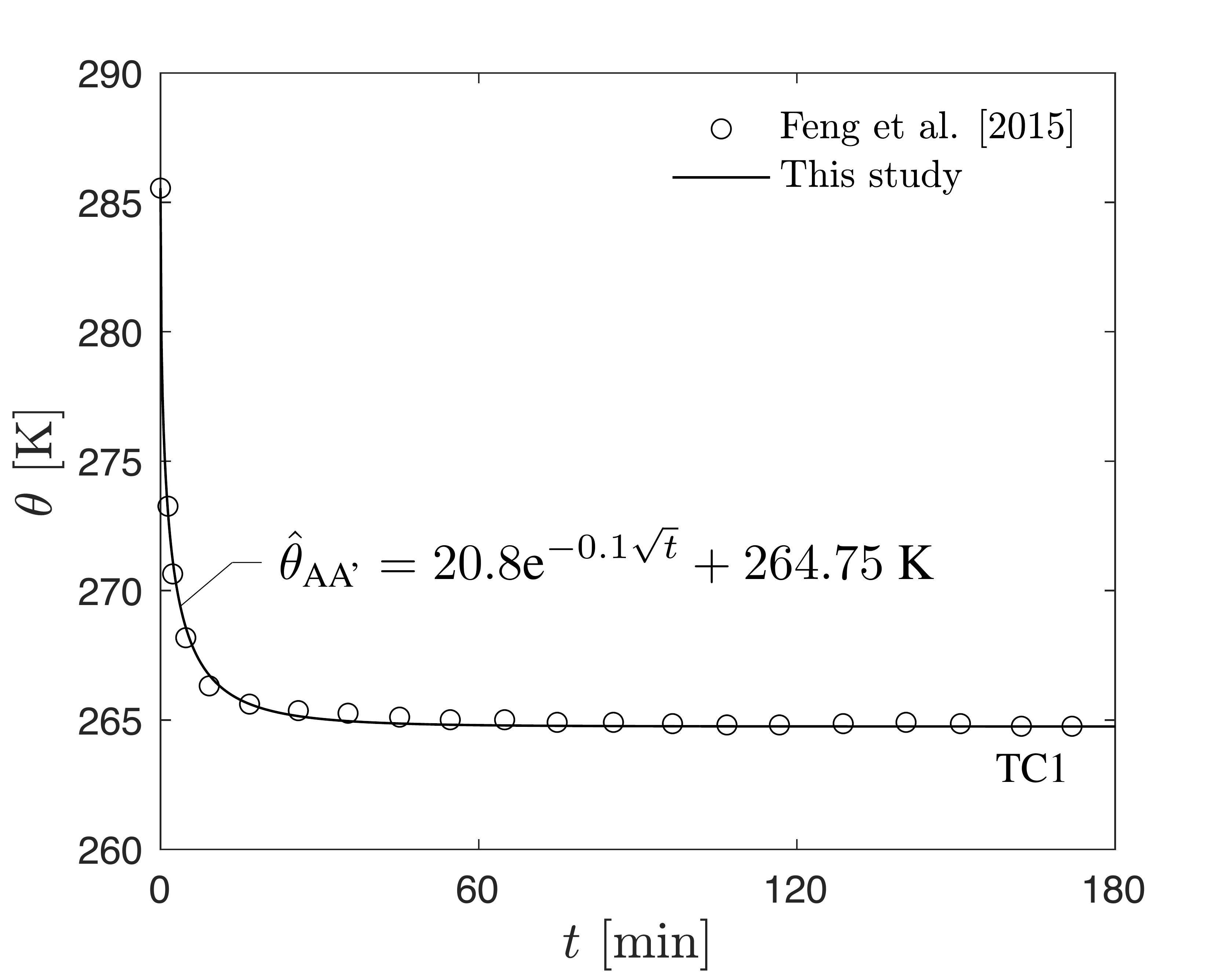}}
\caption{(a) Schematic of the experimental setup for the unidirectional freezing test conducted in \citep{feng2015unidirectional}; (b) Temperature boundary condition applied at the bottom surface of the copper foam (AA') for the numerical simulation.}
\label{fig:feng}
\end{figure}

\subsection{Validation example: homogeneous freezing}
\label{sec:validation}
This section compares the results obtained from the numerical simulation against the physical experiment conducted by Feng et al. \citep{feng2015unidirectional}. 
This experiment is used as a benchmark since it considers the unidirectional freezing of distilled water filled in porous copper foams, which does not involve a fracturing process and yields a clear water-ice boundary layer due to the microstructural attributes of the host matrix. 
As schematically shown in Fig. \ref{fig:feng_sch}, a 30 mm wide, 50 mm long water-saturated copper foam is mounted on a 4 mm thick copper block. 
While the initial temperature is measured to be $\theta_0 = 285.55$ K, the experiment is performed by applying a constant temperature of $\hat{\theta} = 264.15$ K at the bottom part of the copper block at $t = 0$. 
Temperature measurements during the experiment are made by three thermocouples (TC2-TC4) located at 10 mm, 28 mm, and 46 mm from the bottom of the foam (AA'), whereas TC1 records the temperature of the block. 
For the numerical simulation, instead of considering the problem domain as a layered material, we only focus on the water-saturated copper foam and apply time-dependent Dirichlet boundary condition on AA' by using the temperature measured by TC1 [Fig. \ref{fig:feng_applied_T}]. 
We also assume an unlimited water supply from the top surface by imposing $\hat{p}_w = 0$ and applying a fixed boundary condition at the bottom part of the foam. 
Moreover, we consider two different types of copper foams (Foam 1 and Foam 2) with different initial porosity and thermal conductivity [Fig. \ref{fig:feng_sch}]. 
As summarized in Table \ref{tab:mat_prop_feng}, our numerical simulation directly adopts the same thermal properties compared to the physical experiment whereas the solid phase thermal conductivities of the foams are computed based upon the effective properties reported in \citep{feng2015unidirectional}. 
For all other material parameters that are not specified in \citep{feng2015unidirectional}, we choose the properties that resemble those of the water-saturated copper foam. 
In this section, the Allen-Cahn parameters are chosen as: $\nu_c = 0.0001$ m/s, $\gamma_c = 0.65$ J/m$^2$, $\delta_c = 0.0001$ m, and $\epsilon_c = 0.75$ (J/m)$^{1/2}$, while adopting a structured mesh with $h_e = 2.5$ mm and $\Delta t = 60$ sec. 

\begin{table}[h]
\begin{center}
\begin{tabular}{l l l l}
\toprule
Parameter & Description [Unit] & Value & Reference
\\
\midrule
\multicolumn{1}{l}{$\rho_{s}$} 
& \multicolumn{1}{l}{Intrinsic solid mass density [kg/m$^3$]} 
& \multicolumn{1}{l}{7800.0} 
& \multicolumn{1}{l}{-} 
\\
\multicolumn{1}{l}{$\rho_{w}$} 
& \multicolumn{1}{l}{Intrinsic water mass density [kg/m$^3$]} 
& \multicolumn{1}{l}{1000.0} 
& \multicolumn{1}{l}{\citep{feng2015unidirectional}} 
\\
\multicolumn{1}{l}{$\rho_{i}$} 
& \multicolumn{1}{l}{Intrinsic ice mass density [kg/m$^3$]} 
& \multicolumn{1}{l}{920.0} 
& \multicolumn{1}{l}{\citep{feng2015unidirectional}}
\\
\multicolumn{1}{l}{$c_s$}
& \multicolumn{1}{l}{Specific heat of solid [J/kg/K]}  
& \multicolumn{1}{l}{$0.385 \times 10^{3}$}  
& \multicolumn{1}{l}{-}
\\
\multicolumn{1}{l}{$c_w$}
& \multicolumn{1}{l}{Specific heat of water [J/kg/K]}  
& \multicolumn{1}{l}{$4.216 \times 10^{3}$}  
& \multicolumn{1}{l}{\citep{feng2015unidirectional}}
\\
\multicolumn{1}{l}{$c_i$}
& \multicolumn{1}{l}{Specific heat of ice [J/kg/K]}  
& \multicolumn{1}{l}{$2.040 \times 10^{3}$}  
& \multicolumn{1}{l}{\citep{feng2015unidirectional}}
\\
\multicolumn{1}{l}{$\kappa_s$}
& \multicolumn{1}{l}{Thermal conductivity of solid [W/m/K]}  
& \multicolumn{1}{l}{62.855, 44.48}  
& \multicolumn{1}{l}{\citep{feng2015unidirectional}}
\\
\multicolumn{1}{l}{$\kappa_w$}
& \multicolumn{1}{l}{Thermal conductivity of water [W/m/K]}  
& \multicolumn{1}{l}{0.56}  
& \multicolumn{1}{l}{\citep{feng2015unidirectional}}
\\
\multicolumn{1}{l}{$\kappa_i$}
& \multicolumn{1}{l}{Thermal conductivity of ice [W/m/K]}  
& \multicolumn{1}{l}{1.90} 
& \multicolumn{1}{l}{\citep{feng2015unidirectional}}
\\
\multicolumn{1}{l}{$K$}
& \multicolumn{1}{l}{Bulk modulus of solid skeleton [Pa]}  
& \multicolumn{1}{l}{$93.75 \times 10^9$} 
& \multicolumn{1}{l}{-}
\\
\multicolumn{1}{l}{$K_i$}
& \multicolumn{1}{l}{Bulk modulus of ice [Pa]}  
& \multicolumn{1}{l}{$5.56 \times 10^9$} 
& \multicolumn{1}{l}{-}
\\
\multicolumn{1}{l}{$G$}
& \multicolumn{1}{l}{Shear modulus of solid skeleton [Pa]}  
& \multicolumn{1}{l}{$33.58 \times 10^9$} 
& \multicolumn{1}{l}{-}
\\
\multicolumn{1}{l}{$G_i$}
& \multicolumn{1}{l}{Shear modulus of ice [Pa]}  
& \multicolumn{1}{l}{$4.20 \times 10^9$} 
& \multicolumn{1}{l}{-}
\\
\multicolumn{1}{l}{$\phi_0$}
& \multicolumn{1}{l}{Initial porosity [-]}  
& \multicolumn{1}{l}{0.96, 0.98} 
& \multicolumn{1}{l}{\citep{feng2015unidirectional}}
\\
\multicolumn{1}{l}{$k_{\text{mat}}$}
& \multicolumn{1}{l}{Matrix permeability [m$^2$]}  
& \multicolumn{1}{l}{$3.25 \times 10^{-7}$} 
& \multicolumn{1}{l}{-}
\\
\multicolumn{1}{l}{$\mu_w$}
& \multicolumn{1}{l}{Viscosity of water [Pa$\cdot$s]}  
& \multicolumn{1}{l}{$1.0 \times 10^{-3}$} 
& \multicolumn{1}{l}{-}
\\
\multicolumn{1}{l}{$\alpha_{v,\text{int}}$}
& \multicolumn{1}{l}{Volumetric expansion coefficient [-]}  
& \multicolumn{1}{l}{$1.0 \times 10^{-3}$} 
& \multicolumn{1}{l}{-}
\\
\bottomrule
\end{tabular}
\caption{Material parameters for the validation exercise.}
\label{tab:mat_prop_feng}
\end{center}
\end{table}

\begin{figure}[H]
\centering
\includegraphics[height=0.45\textwidth]{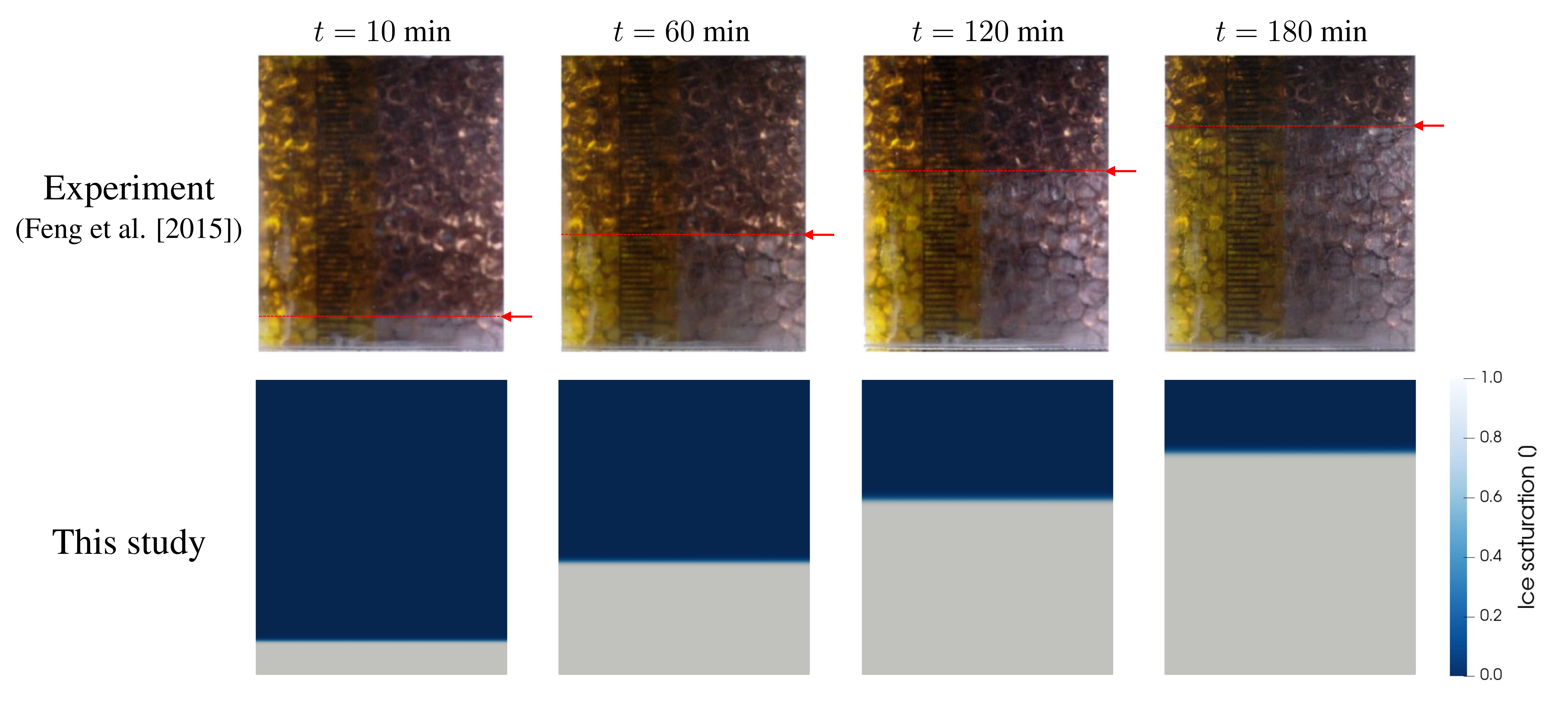}
\caption{Comparison between the physical and numerical experiments on the evolution of the water-ice interface.}
\label{fig:feng_comp}
\end{figure}

\begin{figure}[H]
\centering
\subfigure[]{\label{fig:feng_FF}\includegraphics[height=0.375\textwidth]{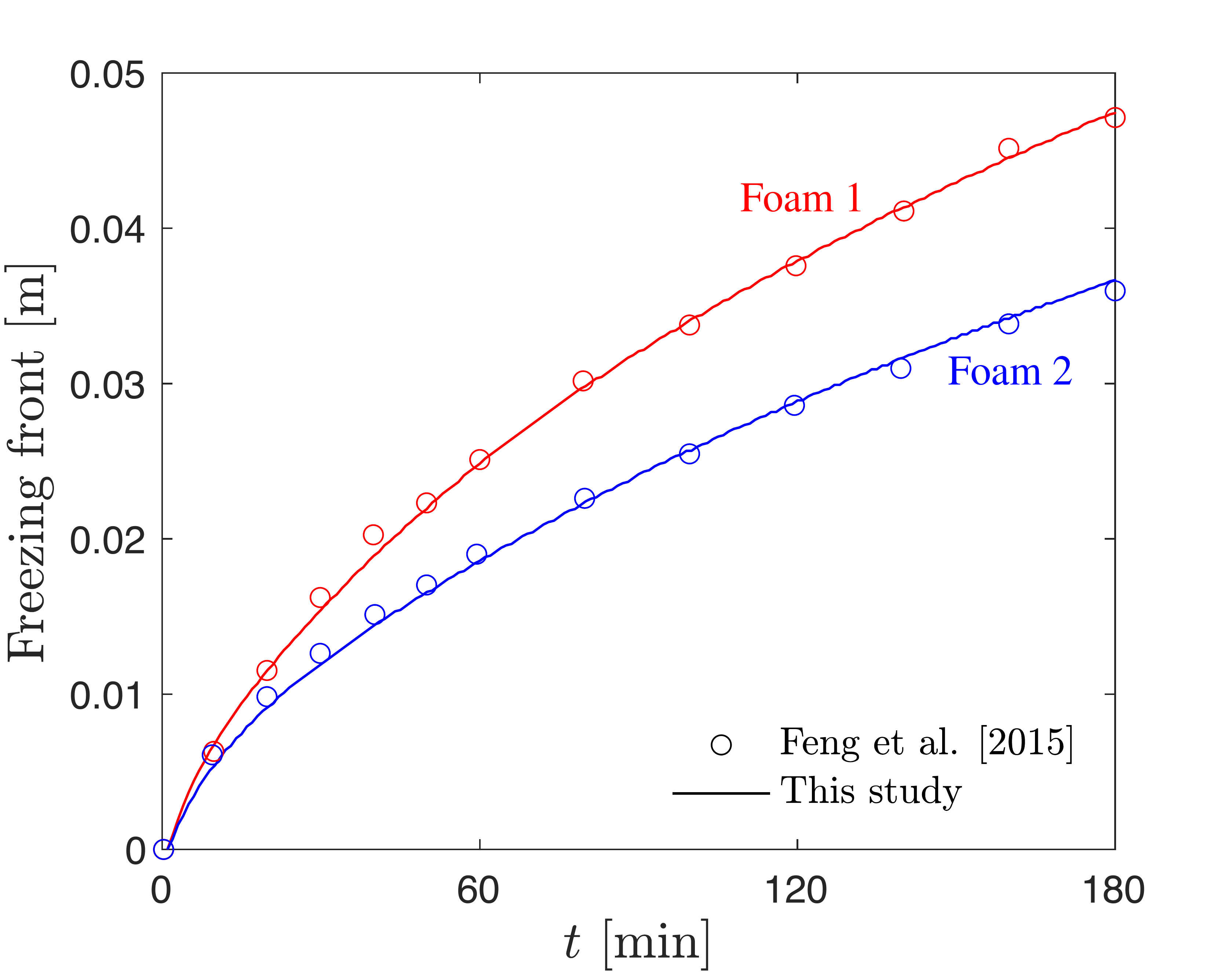}}
\hspace{0.01\textwidth}
\subfigure[]{\label{fig:feng_T}\includegraphics[height=0.375\textwidth]{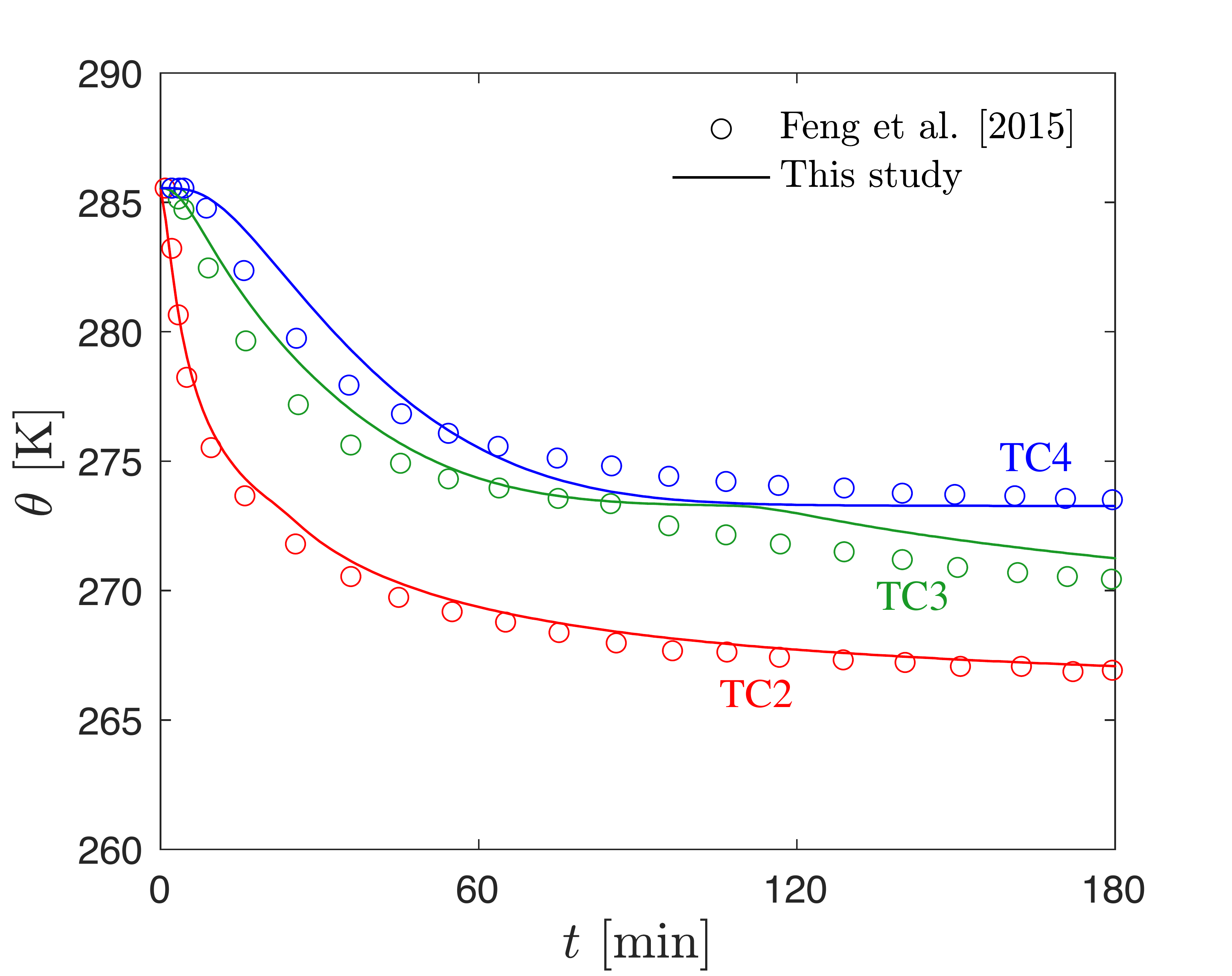}}
\caption{(a) Evolution of the freezing front over time; (b) Temperature variation within Foam 2 measured from TC2, TC3, and TC4.}
\label{fig:feng_results}
\end{figure}

\begin{figure}[H]
\centering
\subfigure[]{\label{fig:feng_cryo}\includegraphics[height=0.375\textwidth]{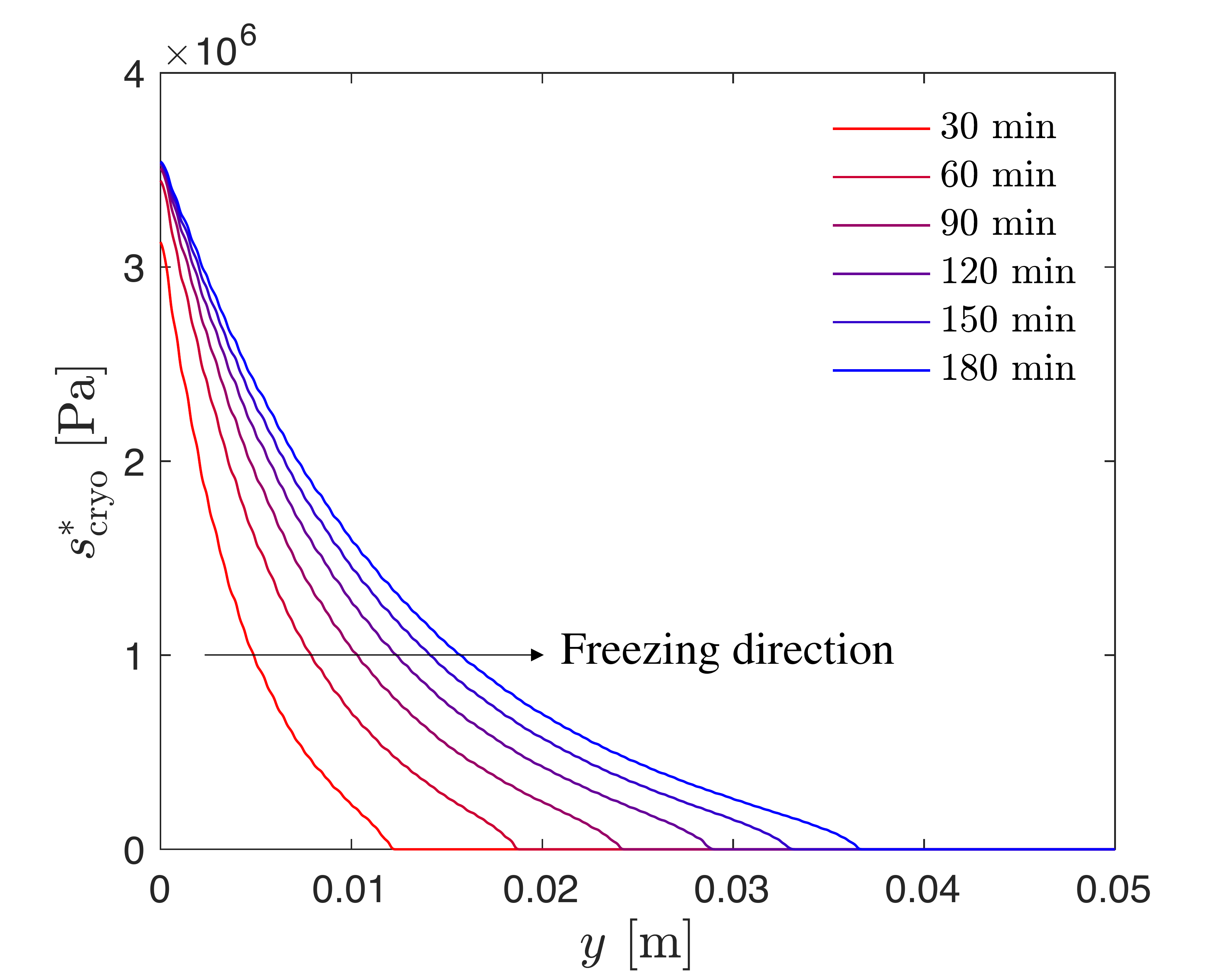}}
\hspace{0.01\textwidth}
\subfigure[]{\label{fig:feng_u}\includegraphics[height=0.375\textwidth]{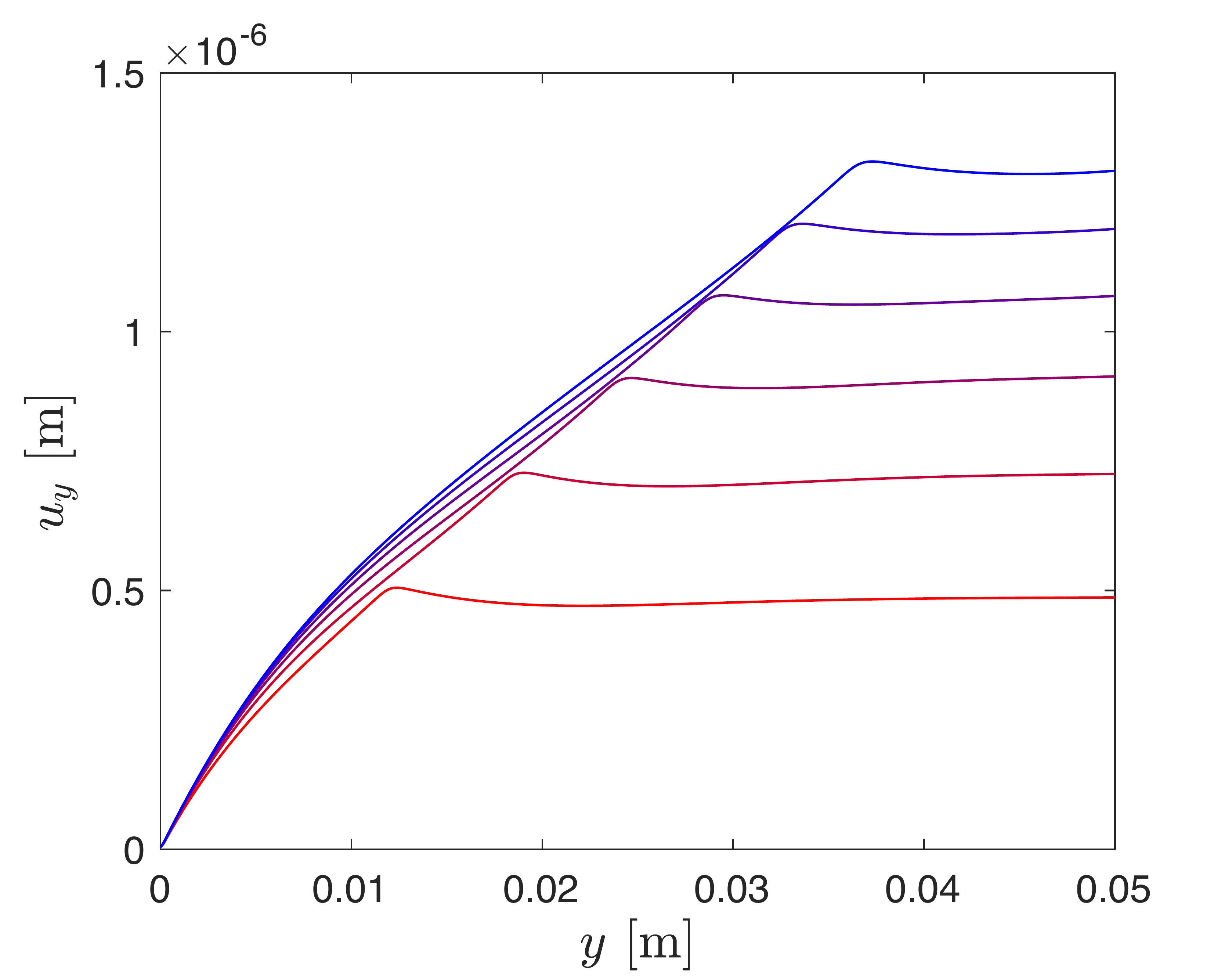}}
\caption{Hydro-mechanical response of Foam 2 subjected to freezing: (a) cryo-suction ($s_{\text{cryo}}^*$) and (b) vertical displacement ($u_y$) profiles.}
\label{fig:feng_results2}
\end{figure}

Fig. \ref{fig:feng_comp} illustrates the evolution of the freezing front within a water-saturated copper foam (Foam 2). 
In both the physical and numerical experiments, water freezing starts from the bottom (AA') and migrates towards the upper part of the foam over time, depending on the conductive heat transfer process. 
While it shows a qualitative agreement between the two, Fig. \ref{fig:feng_results} quantitatively confirms the validity of our model, where we use the circular symbols to indicate the experimental measurements whereas the solid curves denote the numerical results. 
As shown in Fig. \ref{fig:feng_FF}, since Foam 1 possesses higher solidity (lower porosity) compared to Foam 2, the water-ice interface tends to grow relatively faster because it exhibits higher effective thermal conductivity. 
In addition, temperature variations illustrated in Fig. \ref{fig:feng_T} clearly show the interplay between the thermal boundary layer growth and the latent heat, resulting in a nonlinear evolution of the freezing front. 
Although has not been measured experimentally, we further investigate the time-dependent hydro-mechanical response of the specimen from the simulation results shown in Fig. \ref{fig:feng_results2}. 
Based on the freezing retention curve [Eq.~\eqref{eq:freezing_retention}] adopted in this study, positive suction starts to develop if $\theta < \theta_m$ while the region where $s_{\text{cryo}}^* > 0$ evolves over time following the same trajectory of that of the freezing front [Fig. \ref{fig:feng_cryo}]. 
This process also involves a volumetric expansion of the specimen that leads to an increase of the vertical displacement as shown in Fig. \ref{fig:feng_u}, due to the difference between water ($\rho_w$) and ice densities ($\rho_i$). 
It should be also noted that the freezing front always exhibits the largest vertical displacement, implying that the water migration towards the freezing front induced by the suction triggers the consolidation process above the frozen area, resulting in a small volumetric compression therein. 
This observation agrees with the explanation in \citep{amato2021glimpse} where the consolidation front of a frozen soil has been observed experimentally, which corroborates the applicability of our proposed model. 

\subsection{Freeze-thaw action: multiple ice lens growth and thawing in heterogeneous soil}
\label{sec:ice_lens}
\begin{figure}[h]
\centering
\subfigure[]{\label{fig:layer_sch}\includegraphics[height=0.395\textwidth]{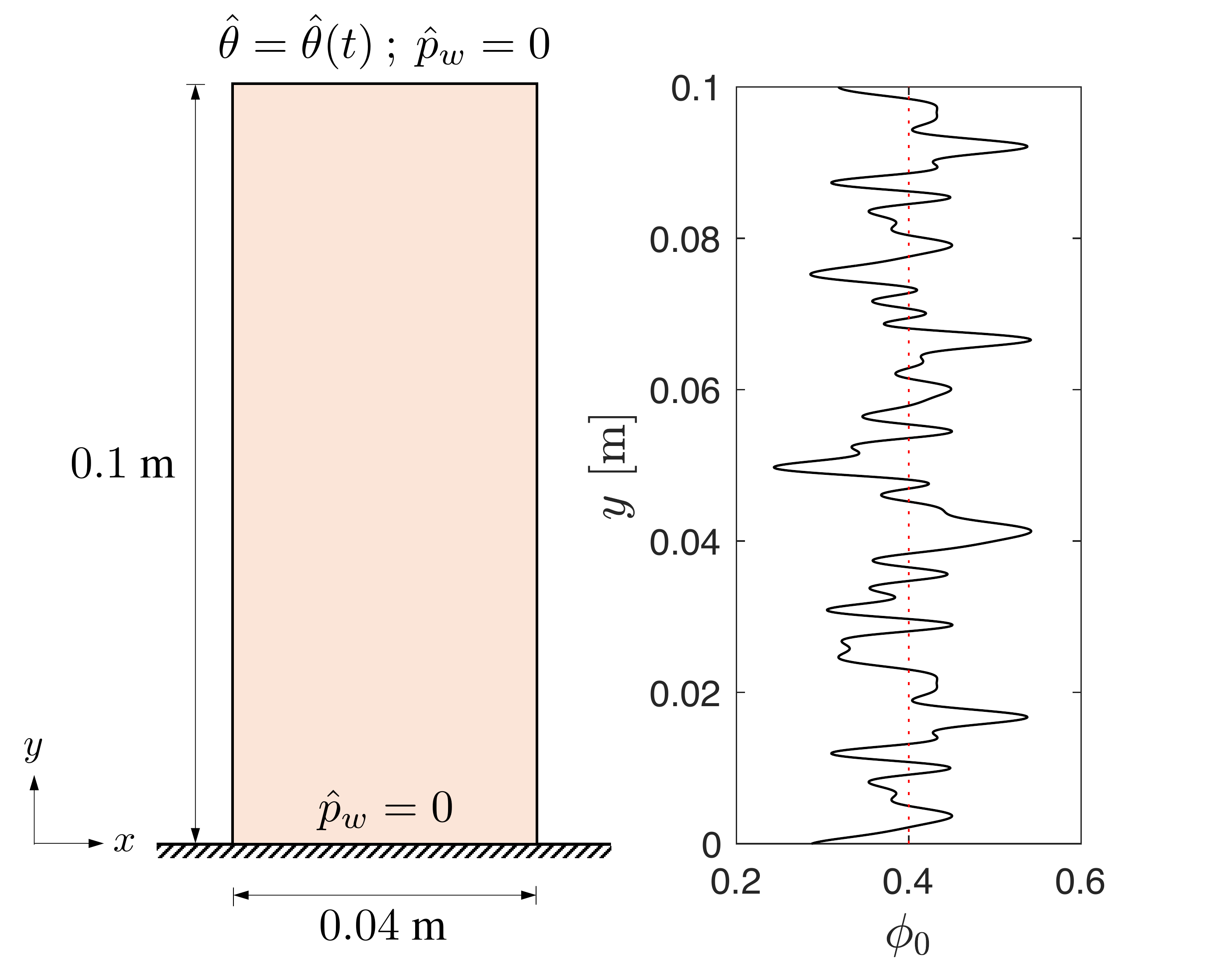}}
\hspace{0.01\textwidth}
\subfigure[]{\label{fig:layer_appT}\includegraphics[height=0.375\textwidth]{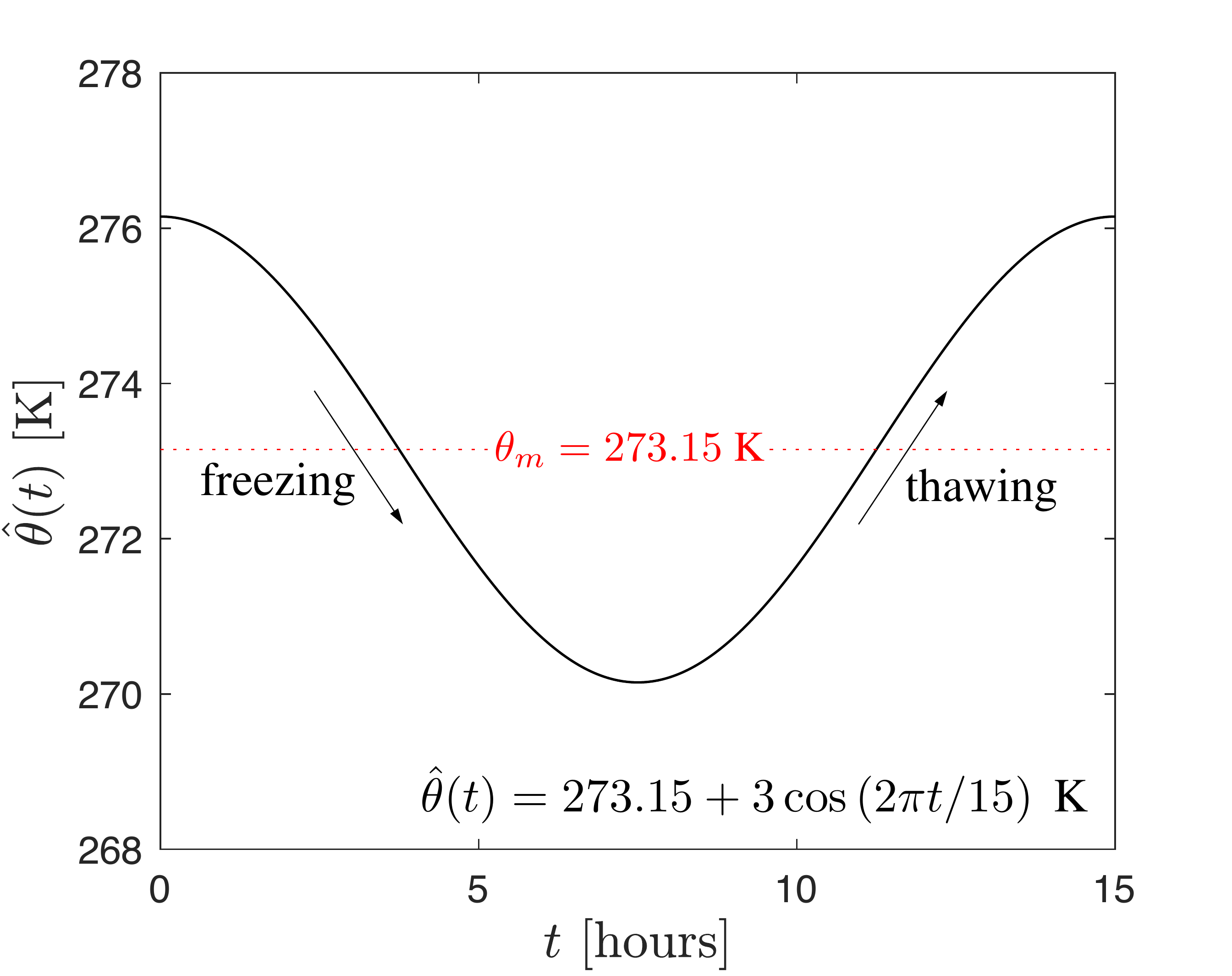}}
\caption{(a) Schematic of geometry and boundary conditions for the numerical freeze-thaw test; (b) Temperature boundary condition applied at the top surface.}
\label{fig:layer1}
\end{figure}

In this section, we showcase the capability of our proposed model by simulating the formation and melting of multiple ice lenses inside a heterogeneous clayey soil specimen. 
As illustrated in Fig. \ref{fig:layer_sch}, the problem domain is 0.04 m wide and 0.1 m long soil column that possesses a random porosity profile along the vertical axis with a mean value of $\phi_{\text{ref}} = 0.4$. 
In addition, we introduce a set of heterogeneous material properties that solely depends on the spatial distribution of initial porosity $\phi_0$. 
Specifically, we adopt a phenomenological model proposed by \citep{uyanik2019estimation} for the shear modulus $G$, while we use a power law for the critical energy release rate $\mathcal{G}_d$ similar to \citep{dunn1973porosity, wang2017unified}:
\begin{equation}
\label{eq:example_G_and_Gc}
G = \frac{3}{2} \left( \frac{1 - 2 \nu}{1 + \nu} \right) \exp{\left[ 10 (1 - \phi_0)\right]} \text{ [MPa]}
\: \: ; \: \:
\mathcal{G}_d = \mathcal{G}_{d,\text{ref}} \left( \frac{1 - \phi_0}{1 - \phi_{\text{ref}}} \right)^{n_{\phi}}.
\end{equation}
Here, we assume that the Poisson's ratio remains constant $\nu = 0.25$ throughout the entire domain while we set $\mathcal{G}_{d,\text{ref}} = 4.5$ N/m and $n_{\phi} = 10$. 
For all other material properties that are homogeneous, as summarized in Table \ref{tab:mat_prop_layer}, we choose values similar to those of the clayey soil. 
It should be noted that we adopt $\alpha_{v,\text{dam}} = 0.08$ which is identical to the theoretical value of $1 - \rho_i/\rho_w$ for the expansion coefficient, whereas we set $\alpha_{v,\text{int}} = 0.01 \alpha_{v,\text{dam}}$ due to the existence of thin water film between the intact solid and the pore ice. 
Meanwhile, the parameters for the Allen-Cahn phase field equation are chosen as: $\nu_c = 0.0001$ m/s, $\gamma_c = 0.65$ J/m$^2$, $\delta_c = 0.0001$ m, and $\epsilon_c = 1.0$ (J/m)$^{1/2}$, whereas we set $h_e = 0.5$ mm and $\Delta t = 60$ sec. 

\begin{table}[h]
\begin{center}
\begin{tabular}{l l l}
\toprule
Parameter & Description [Unit] & Value
\\
\midrule
\multicolumn{1}{l}{$\rho_{s}$} 
& \multicolumn{1}{l}{Intrinsic solid mass density [kg/m$^3$]} 
& \multicolumn{1}{l}{2650.0} 
\\
\multicolumn{1}{l}{$\rho_{w}$} 
& \multicolumn{1}{l}{Intrinsic water mass density [kg/m$^3$]} 
& \multicolumn{1}{l}{1000.0} 
\\
\multicolumn{1}{l}{$\rho_{i}$} 
& \multicolumn{1}{l}{Intrinsic ice mass density [kg/m$^3$]} 
& \multicolumn{1}{l}{920.0} 
\\
\multicolumn{1}{l}{$c_s$}
& \multicolumn{1}{l}{Specific heat of solid [J/kg/K]}  
& \multicolumn{1}{l}{$0.75 \times 10^{3}$}  
\\
\multicolumn{1}{l}{$c_w$}
& \multicolumn{1}{l}{Specific heat of water [J/kg/K]}  
& \multicolumn{1}{l}{$4.20 \times 10^{3}$}  
\\
\multicolumn{1}{l}{$c_i$}
& \multicolumn{1}{l}{Specific heat of ice [J/kg/K]}  
& \multicolumn{1}{l}{$1.90 \times 10^{3}$}  
\\
\multicolumn{1}{l}{$\kappa_s$}
& \multicolumn{1}{l}{Thermal conductivity of solid [W/m/K]}  
& \multicolumn{1}{l}{7.69}  
\\
\multicolumn{1}{l}{$\kappa_w$}
& \multicolumn{1}{l}{Thermal conductivity of water [W/m/K]}  
& \multicolumn{1}{l}{0.56}  
\\
\multicolumn{1}{l}{$\kappa_i$}
& \multicolumn{1}{l}{Thermal conductivity of ice [W/m/K]}  
& \multicolumn{1}{l}{2.25} 
\\
\multicolumn{1}{l}{$K_i$}
& \multicolumn{1}{l}{Bulk modulus of ice [Pa]}  
& \multicolumn{1}{l}{$5.56 \times 10^9$} 
\\
\multicolumn{1}{l}{$G_i$}
& \multicolumn{1}{l}{Shear modulus of ice [Pa]}  
& \multicolumn{1}{l}{$4.20 \times 10^9$} 
\\
\multicolumn{1}{l}{$\phi_{\text{ref}}$}
& \multicolumn{1}{l}{Reference porosity [-]}  
& \multicolumn{1}{l}{0.4} 
\\
\multicolumn{1}{l}{$k_{\text{mat}}$}
& \multicolumn{1}{l}{Matrix permeability [m$^2$]}  
& \multicolumn{1}{l}{$1.0 \times 10^{-13}$} 
\\
\multicolumn{1}{l}{$\mu_w$}
& \multicolumn{1}{l}{Viscosity of water [Pa$\cdot$s]}  
& \multicolumn{1}{l}{$1.0 \times 10^{-3}$} 
\\
\multicolumn{1}{l}{$\mathcal{G}_{d,\text{ref}}$}
& \multicolumn{1}{l}{Reference critical energy release rate [N/m]}  
& \multicolumn{1}{l}{$4.5$} 
\\
\multicolumn{1}{l}{$l_d$}
& \multicolumn{1}{l}{Regularization length scale parameter [m]}  
& \multicolumn{1}{l}{$1.0 \times 10^{-3}$} 
\\
\multicolumn{1}{l}{$\mathcal{H}_{\text{crit}}$}
& \multicolumn{1}{l}{Normalized threshold strain energy [-]}  
& \multicolumn{1}{l}{$0.05$} 
\\
\multicolumn{1}{l}{$\alpha_{v,\text{int}}$}
& \multicolumn{1}{l}{Volumetric expansion coefficient (intact) [-]}  
& \multicolumn{1}{l}{$0.8 \times 10^{-3}$} 
\\
\multicolumn{1}{l}{$\alpha_{v,\text{dam}}$}
& \multicolumn{1}{l}{Volumetric expansion coefficient (damaged) [-]}  
& \multicolumn{1}{l}{$80.0 \times 10^{-3}$} 
\\
\multicolumn{1}{l}{$K_c^*$}
& \multicolumn{1}{l}{Kinetic parameter [Pa]}  
& \multicolumn{1}{l}{$5.0 \times 10^9$} 
\\
\multicolumn{1}{l}{$g_c^*$}
& \multicolumn{1}{l}{Kinetic parameter [-]}  
& \multicolumn{1}{l}{$1.25$} 
\\
\bottomrule
\end{tabular}
\caption{Material parameters for the numerical freeze-thaw test.}
\label{tab:mat_prop_layer}
\end{center}
\end{table}

While we set the initial temperature as $\theta_0 = 276.15$ K, the numerical freeze-thaw test is performed by applying a time-dependent temperature boundary condition at the top, represented by a sinusoidal function. 
As shown in Fig. \ref{fig:layer_appT}, the freezing process starts at $t = 3.75$ hr and continues until the top surface temperature reaches the melting temperature of $\theta_m = 273.15$ K at $t = 11.25$ hr, where the frozen soil begins to thaw. 
During the simulation, the bottom part of the specimen is held fixed while we prescribe zero pore water pressure boundaries ($\hat{p}_w = 0$) at both the top and the bottom surfaces. 
The left and right boundaries, on the other hand, are subjected to zero water mass flux and heat flux conditions. 
Based on this setting, the water is supplied from the bottom during the freezing phase, while the water expulsion towards the top surface during the melting phase leads to a thawing settlement of the specimen. 

\begin{figure}[h]
\centering
\includegraphics[height=0.35\textwidth]{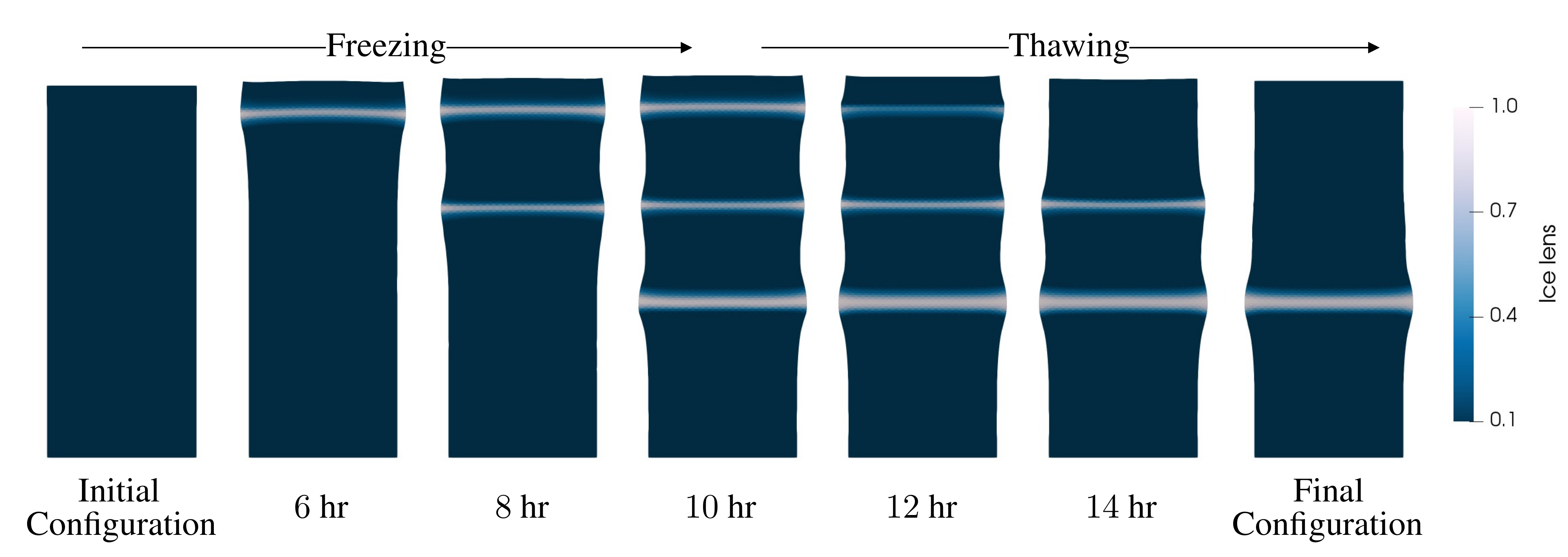}
\caption{Formation and melting of multiple ice lenses during the numerical freeze-thaw test.}
\label{fig:layer_lens_config}
\end{figure}

\begin{figure}[h]
\centering
\subfigure[]{\label{fig:freezing_uy}\includegraphics[height=0.35\textwidth]{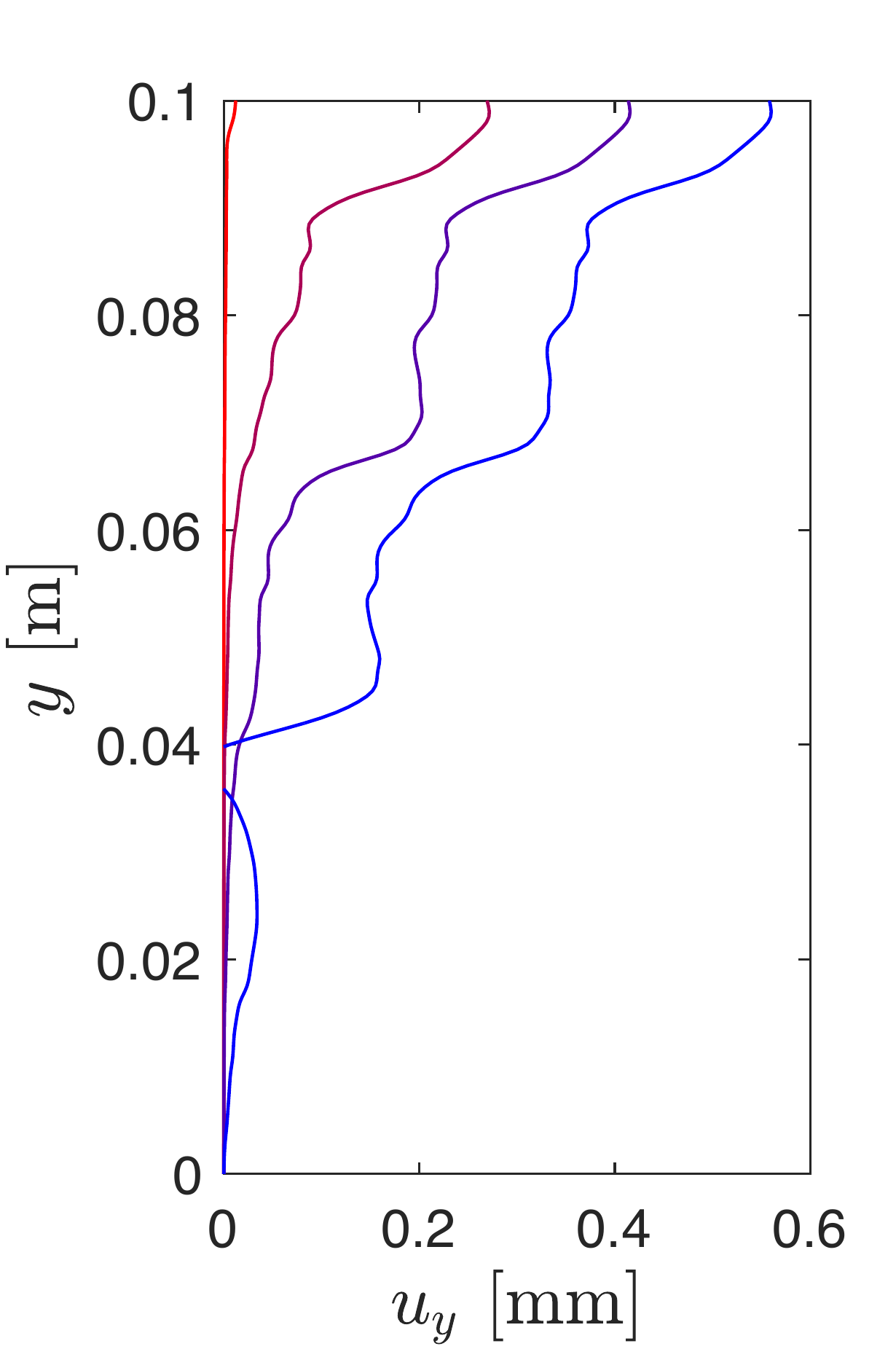}}
\hspace{0.01\textwidth}
\subfigure[]{\label{fig:freezing_scryo}\includegraphics[height=0.35\textwidth]{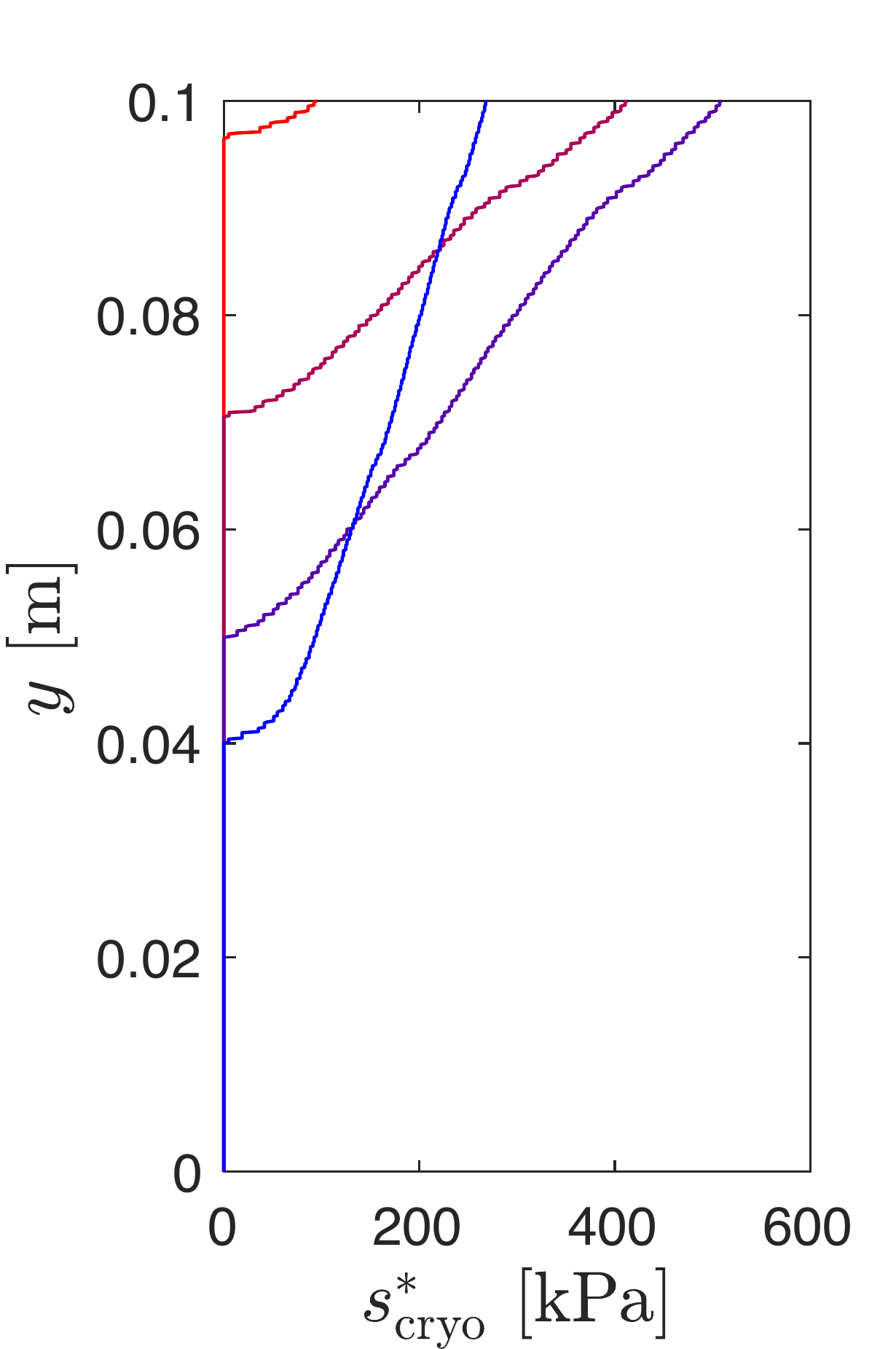}}
\hspace{0.01\textwidth}
\subfigure[]{\label{fig:freezing_theta}\includegraphics[height=0.35\textwidth]{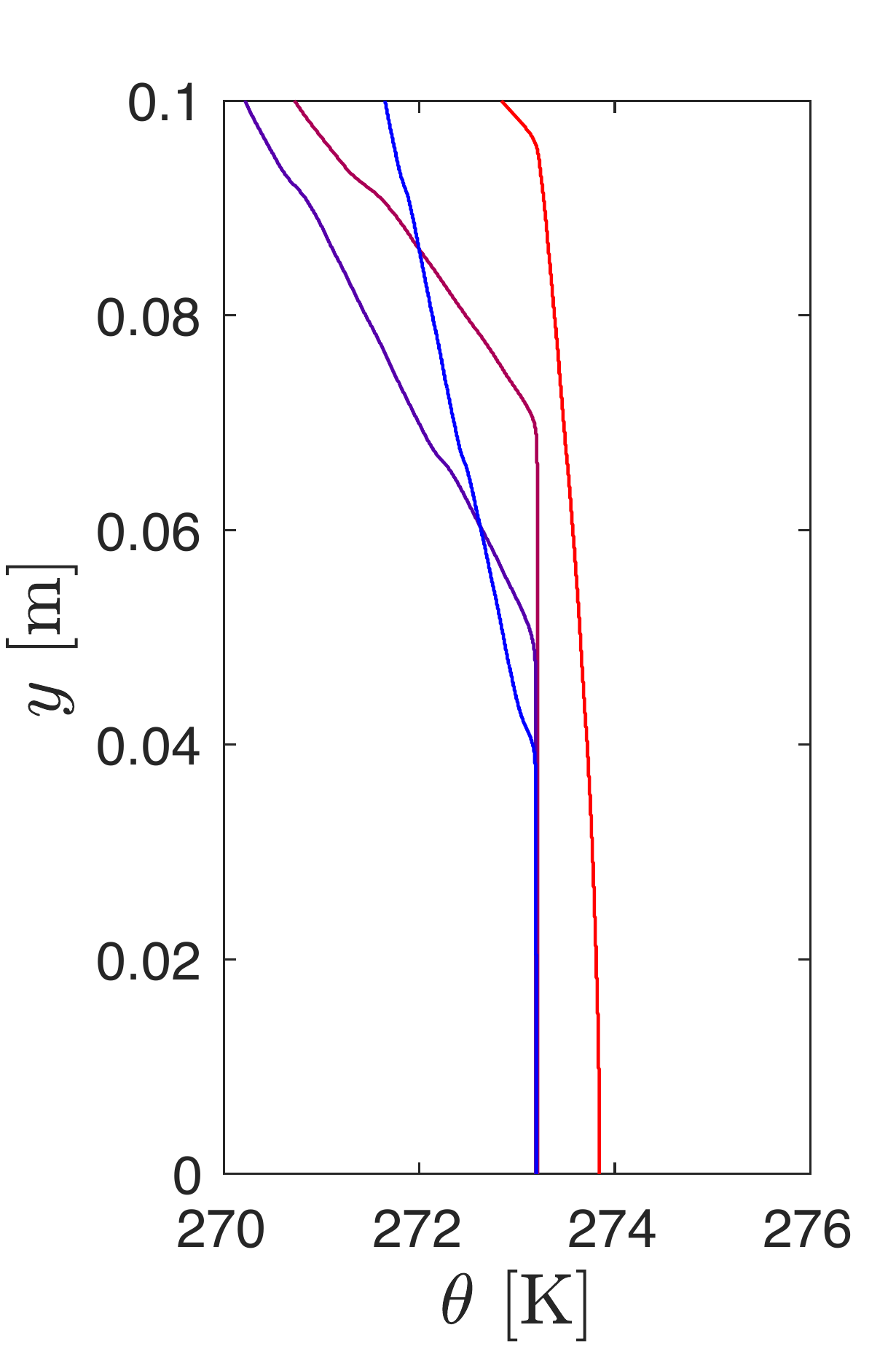}}
\hspace{0.01\textwidth}
\subfigure[]{\label{fig:freezing_Si}\includegraphics[height=0.35\textwidth]{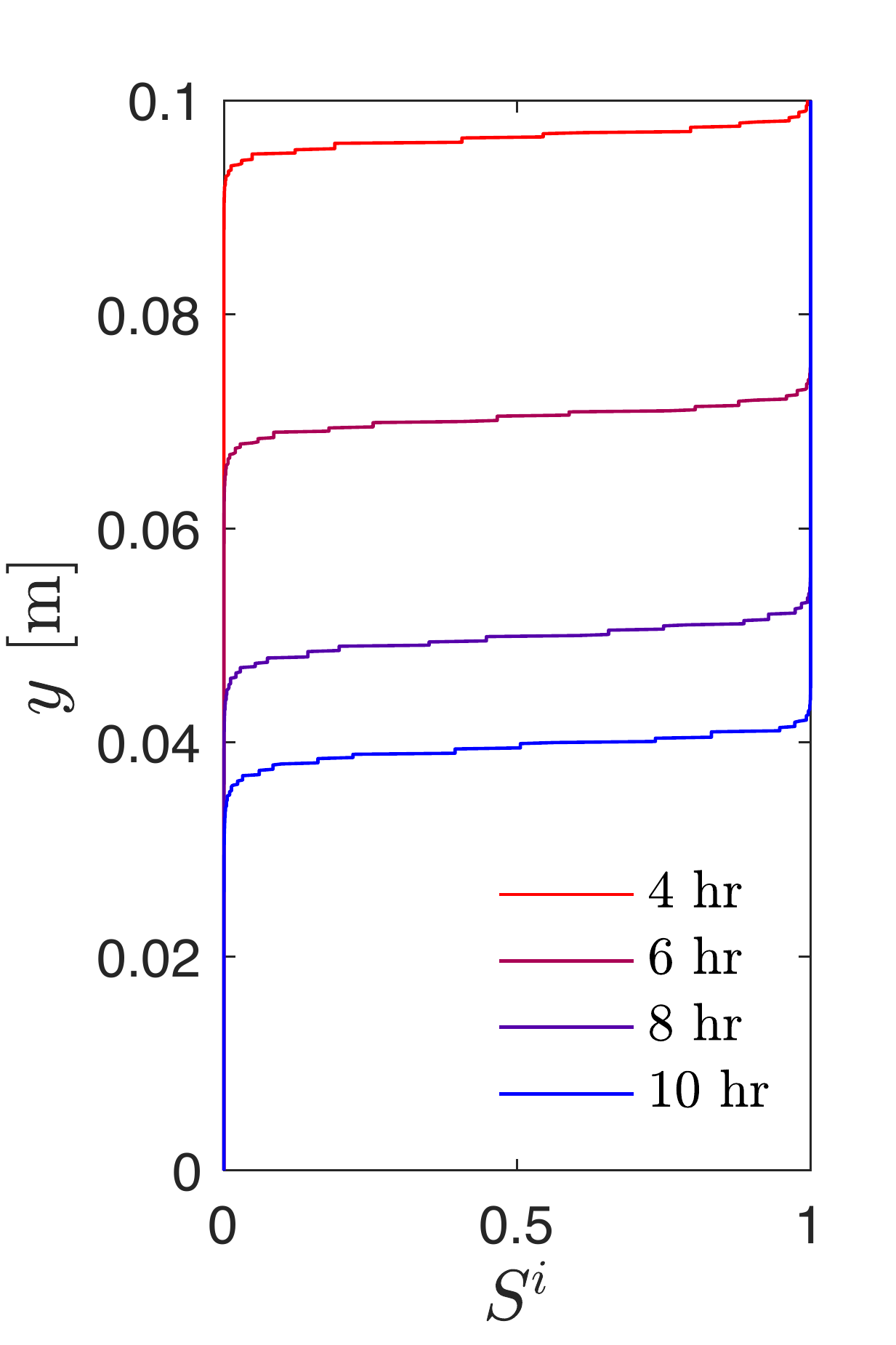}}
\caption{Thermo-hydro-mechanical response of the specimen during the freezing phase: (a) vertical displacement ($u_y$), (b) cryo-suction ($s_{\text{cryo}}^*$), (c) temperature ($\theta$), and (d) ice saturation ($S^i$) profiles along the central axis.}
\label{fig:layer_freezing}
\end{figure}

Fig. \ref{fig:layer_lens_config} shows the formation and melting of multiple ice lenses during the numerical freeze-thaw test. 
Here, we use a scaling factor of 5 while the color bar represents the value of the indicator function $\chi^i$ defined in Eq.~\eqref{eq:ice_lens_indicator}. 
As illustrated in Fig. \ref{fig:layer_freezing}, the water freezes from the top to the bottom during the freezing phase ($3.75 \text{ hr} \le t \le 11.25 \text{ hr}$), which leads to development of the cryo-suction and a volumetric expansion due to the phase transition. 
Since the applied temperature at the top starts to increase after reaching its minimum, $s_{\text{cryo}}^*$ tends to decrease after $t = 7.5$ hr due to the freezing characteristic function in Eq.~\eqref{eq:freezing_retention} although the freezing front still propagates towards the bottom. 
Also, during the freezing phase, soil specimen tends to exhibit a constant temperature distribution at the region below the freezing front due to the latent heat effect, similar to our previous example shown in Section \ref{sec:validation}. 
More importantly, we observe a sequential development of the ice lenses at $y = 0.092$ m, $y = 0.066$ m, and $y = 0.042$ m, respectively. 
This result is expected, since those regions possess relatively high initial porosity compared to the other regions [Fig. \ref{fig:layer_sch}]. 
If the freezing front reaches the porous zone where the critical energy release rate is relatively low, both the cryo-suction and the exerted stress due to the phase transition initiate the horizontal crack. 

\begin{figure}[h]
\centering
\subfigure[]{\label{fig:thawing_uy}\includegraphics[height=0.35\textwidth]{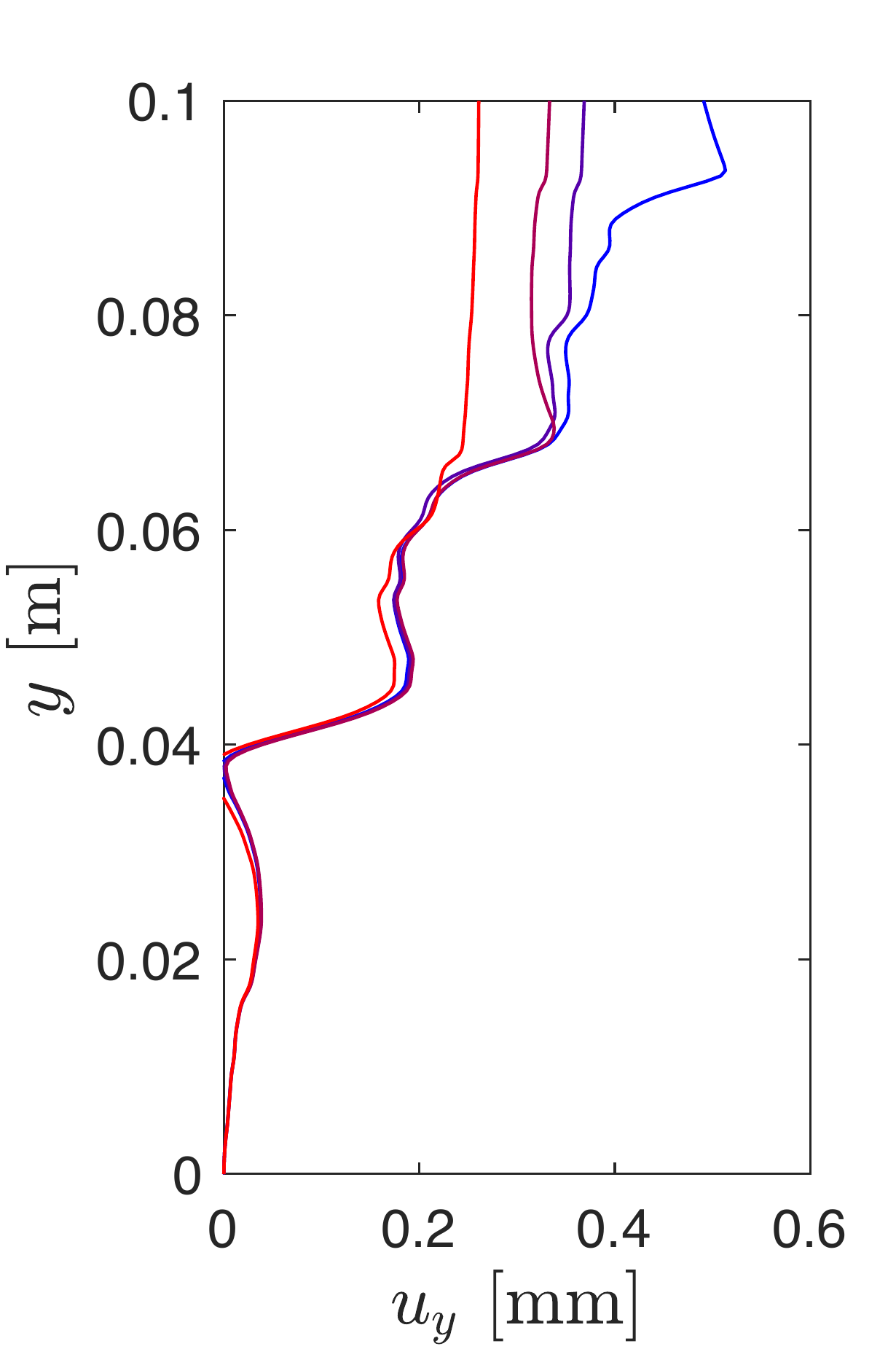}}
\hspace{0.01\textwidth}
\subfigure[]{\label{fig:thawing_scryo}\includegraphics[height=0.35\textwidth]{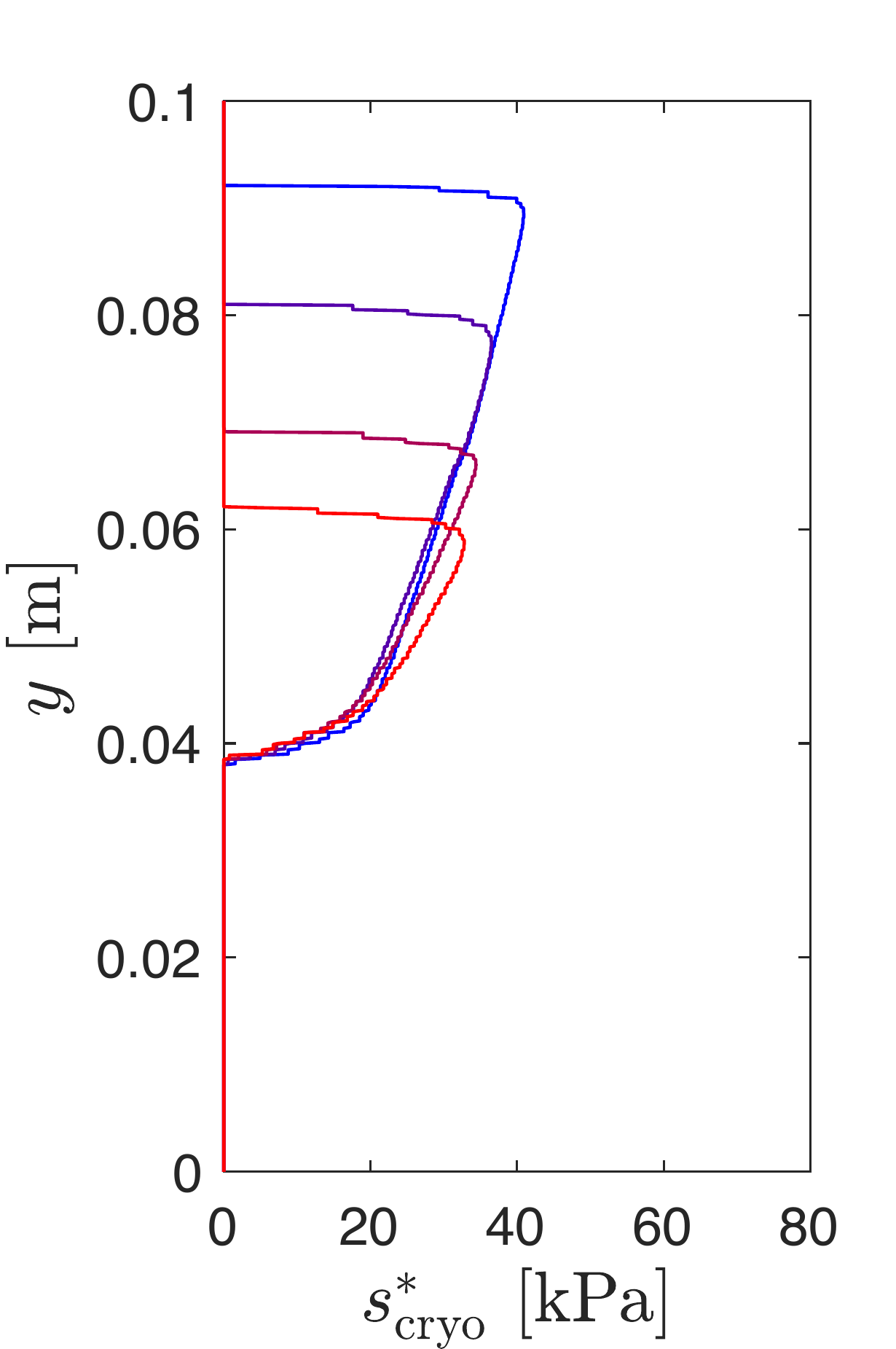}}
\hspace{0.01\textwidth}
\subfigure[]{\label{fig:thawing_theta}\includegraphics[height=0.35\textwidth]{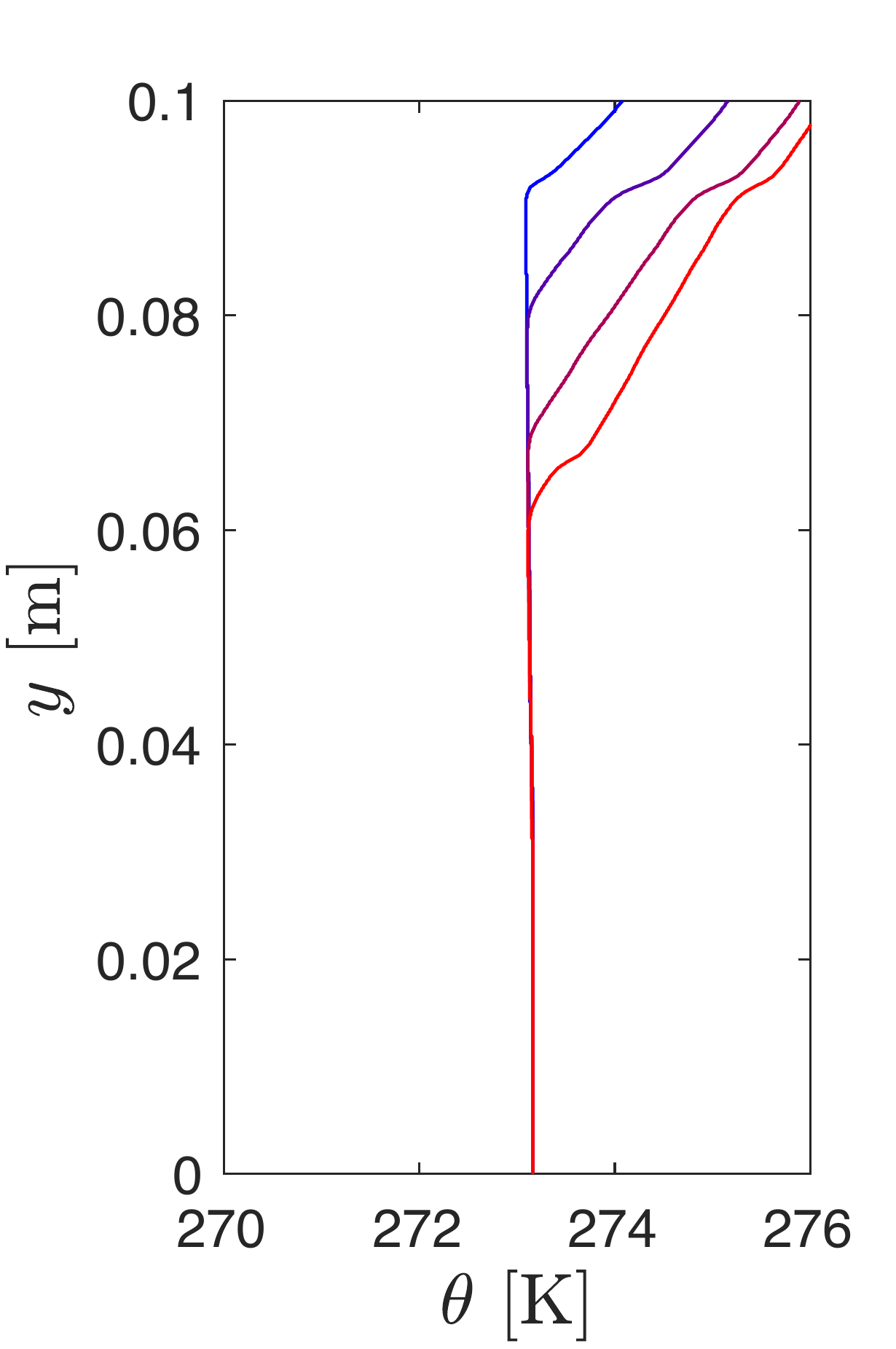}}
\hspace{0.01\textwidth}
\subfigure[]{\label{fig:thawing_Si}\includegraphics[height=0.35\textwidth]{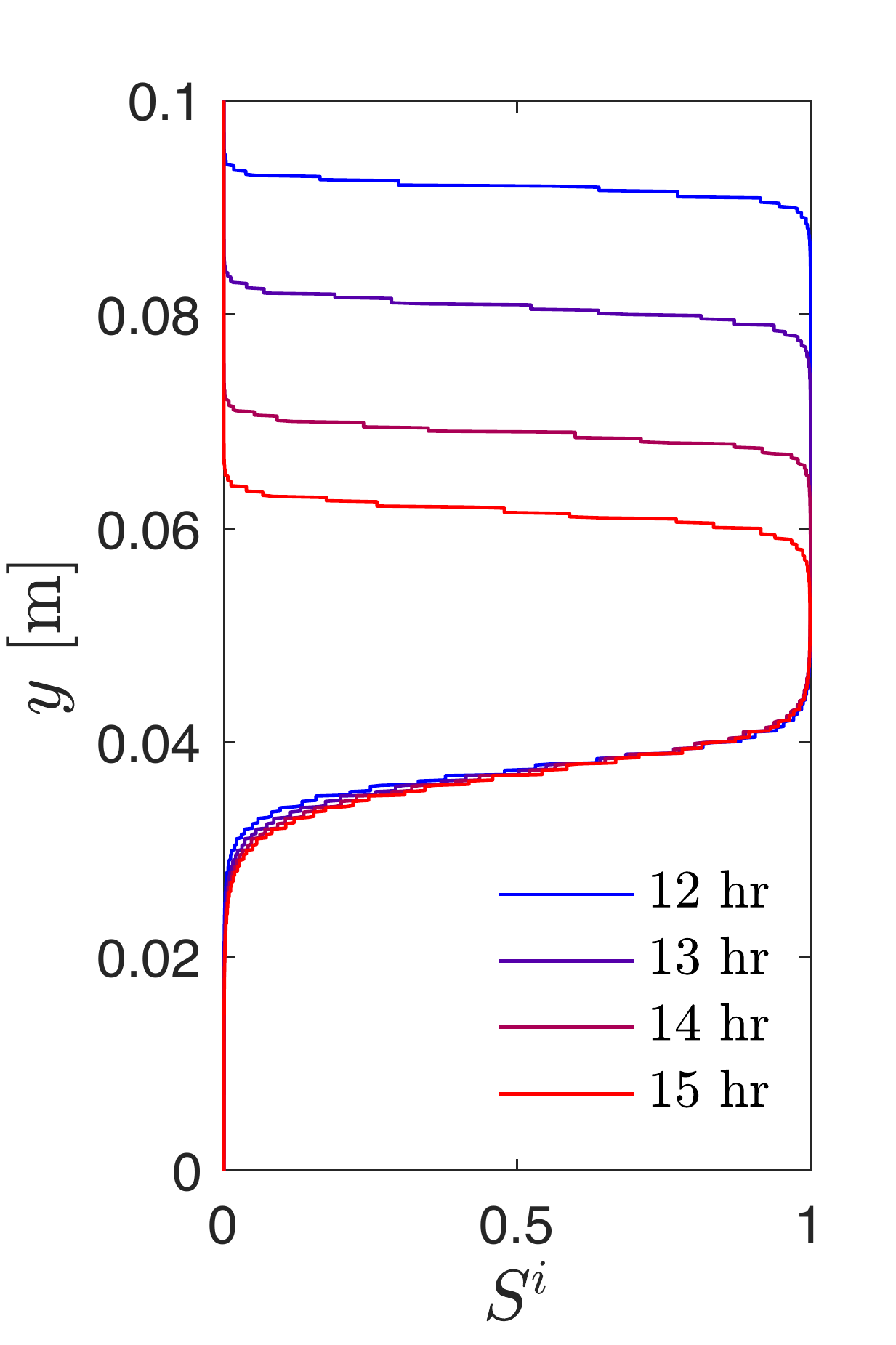}}
\caption{Thermo-hydro-mechanical response of the specimen during the thawing phase: (a) vertical displacement ($u_y$), (b) cryo-suction ($s_{\text{cryo}}^*$), (c) temperature ($\theta$), and (d) ice saturation ($S^i$) profiles along the central axis.}
\label{fig:layer_thawing}
\end{figure}

Once the freezing-induced fracture is developed, segregated bulk ice tends to form inside the opened crack at higher growth rates that lead to an abrupt volume expansion therein (Fig. \ref{fig:layer_lens_config}). 
As illustrated in Fig. \ref{fig:layer_thawing}, we observe the opposite response during the thawing phase ($11.25 \text{ hr} \le t \le 15 \text{ hr}$). 
At $t = 11.25$ hr, once the applied temperature at the top again reaches the melting temperature $\theta_m = 273.15$ K, the soil specimen stops freezing and begins to thaw from the top to the bottom. 
During the thawing process, the melting front tends to move downwards whereas the freezing front remains unchanged since the bottom surface is thermally insulated. 
As the melted region where $\theta > \theta_m$ evolves, the vertical displacement tends to decrease over time due to both the volume contraction during the phase transition and the water expulsion towards the top surface. 

\begin{figure}[H]
\centering
\includegraphics[height=0.375\textwidth]{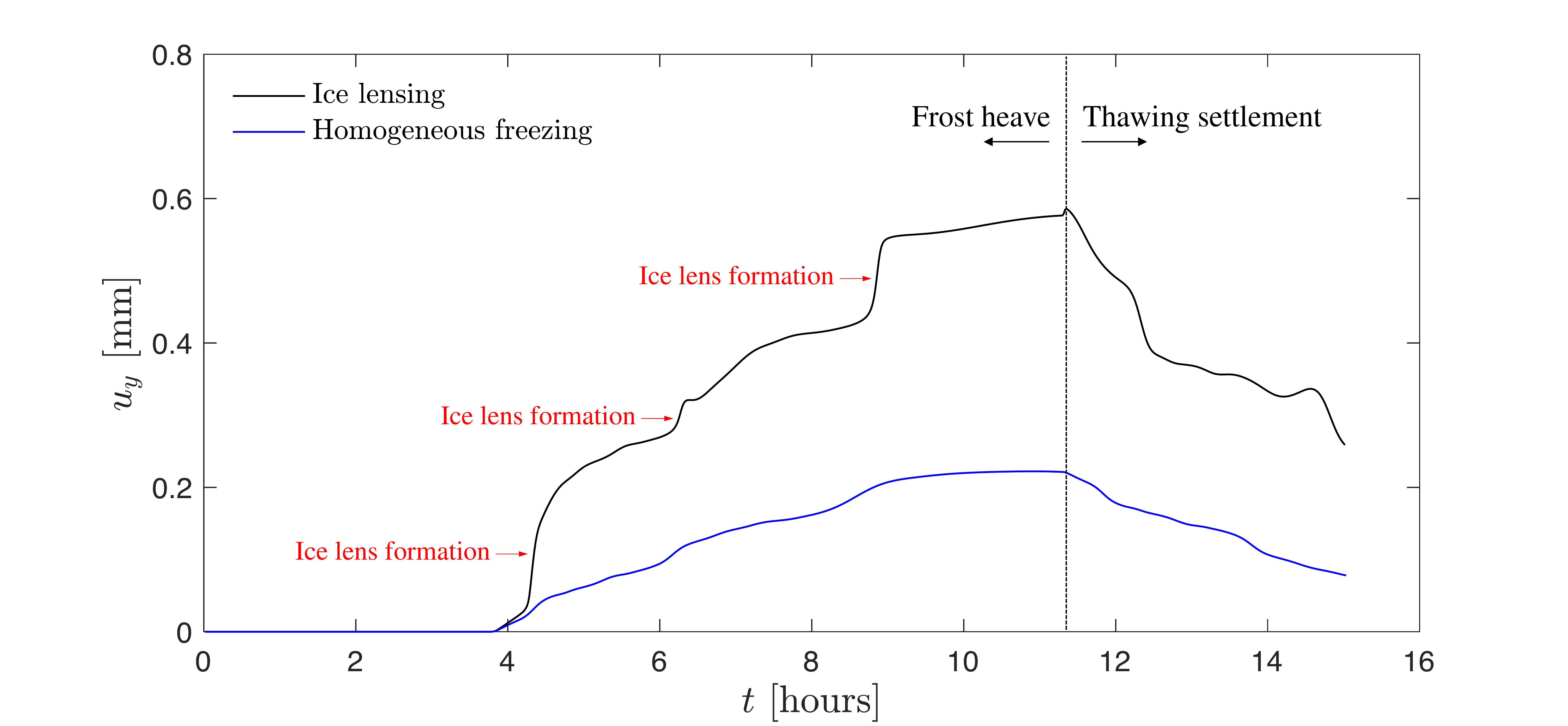}
\caption{Vertical displacement ($u_y$) evolution of the top surface during the numerical freeze-thaw test. The black curve is obtained from a thermo-hydro-mechanical simulation that enables ice lensing; the blue curve is obtained from control experiment that takes out the ice lensing capacity.}
\label{fig:layer_lens_displacement}
\end{figure}

Fig. \ref{fig:layer_lens_displacement} shows the evolution of the vertical displacement of the top surface during the freeze-thaw test (black curve). 
For comparison, we introduce a control experiment where the phase field solvers for both ice lens and damage are disabled but otherwise the material parameters are identical (blue curve). Hence, the numerical specimen in the control experiment may exhibit the homogeneous freezing and thawing but not ice lens formation and melting. 
The frost heave and thawing settlement for both experiments are compared to assess  the impact of the ice lenses 
on the material responses. 

In the prime numerical experiment, ice lenses sequentially develop at $y = 0.092$ m, $y = 0.066$ m, and $y = 0.042$ m (see Fig. \ref{fig:layer_lens_config}), respectively. Each time the ice lens begin leads to expansion, the soil expand more rapidly and hence the steeper slope of the black curve, which indicates rapid expansion of the numerical specimen, at $t = 4.2$ hr, $t = 6.2$ hr, and $t = 8.8$ hr.
During the thawing phase, the prescribed temperature of the top surface increase. This temperature increases leads to abrupt settlement within the first 2 hours of the thawing phase. As the ice lenses melt and subsequently drain out from the domain, the numerical specimen shrinks (black curve). 
In contract, homogeneous freezing and thawing results in considerably less amount of frost heaving and thawing settlement, due to the absence of cracks where ice lenses may form. 

The significant difference between the two simulations have important practical implications. 
It is presumably possible to use optimization algorithm to identify the material parameters such that the control experiment may match better with the observed frost heave and thawing settlements of soil vulnerable to ice lens formation. 
However, the apparent match obtained from such an excessive calibration is fruitless as it may lead to material parameters that are not physics and therefore lead to a model weak at forward predictions. 
Results of these numerical experiments again suggest that the ice lenses play a key role in frost heaving and the subsequent settlement of soils.
This example also highlights that our proposed model is capable of simulating the ice lens growth and thaw in a fluid-saturated porous media, which may not be easily captured via a classical thermo-hydro-mechanical model.

\section{Conclusions}
\label{sec:conclusion}
In this work, we introduce a multi-phase-field microporomechanics theory and the corresponding finite element solver to capture the freeze-thaw action in a frozen/freezing/thawing porous medium that may form ice lenses. 
By introducing two phase field variables that indicate the phase of the ice/water and damaged/undamaged material state, the proposed thermo-hydro-mechanical model is capable of simulating the freezing-induced fracture caused by the growth of the ice lens as segregated ice. 
We also extend the Bishop's effective stress principle for frozen soil to incorporate the effects of damage and ice growth and distinguish them from those of the freezing retention responses. 
This treatment enables us to take into account the shear strength of the ice lenses and analyzes how the homogeneous freezing process and the ice lens growth
affect the thermo-hydro-mechanical coupling effects in the transient regime. 
The model is validated against published freezing experiments. 
To investigate how the formation and thawing of ice lens affect the frost heave and thaw settlement, we conduct numerical experiments that simulate the climate-induced frozen heave and thaw settlement in one thermal cycle and compare the simulation results with those obtained from a thermo-hydro-mechanical model that does not explicitly capture the ice lens. 
The simulation results suggest that explicitly capturing the growth and thaw of ice lens may provide more precise predictions and analyses on the multi-physical coupling effects of frozen soil at different time scales. 
Accurate and precise predictions on the frozen heave and thaw settlement are crucial for many modern engineering applications, from estimating the durability of pavement system to the exploration of ice-rich portions of Mars. 
This work provides a foundation for a more precise depiction of frozen soil by incorporating freezing retention, heat transfer, fluid diffusion, fracture mechanics, and ice lens growth in a single model. 
More accurate predictions nevertheless may require sufficient data to solve the inverse problems and quantify uncertainties as well as optimization techniques to identify material parameters from different experiments. 
Such endeavors are important and will be considered in future study.

\section{Acknowledgments}
\label{ch:acknowledgments}
This work is primarily supported by the Earth Materials and Processes program from the US Army Research Office under grant contract W911NF-18-2-0306, with additional time of the PI supported by the NSF CAREER grant from Mechanics of Materials and Structures program at National Science Foundation under grant contract CMMI-1846875. 
These supports are gratefully acknowledged. 
The views and conclusions contained in this document are those of the authors, and should not be interpreted as representing the official policies, either expressed or implied, of the sponsors, including the Army Research Laboratory or the U.S. Government. 
The U.S. Government is authorized to reproduce and distribute reprints for Government purposes notwithstanding any copyright notation herein.

\bibliographystyle{plainnat}
\bibliography{main}

\end{document}